\documentclass[twocolumn,english,aps,prl,manuscript,reprint,superscriptaddress,showpacs,showkeys,longbibliography]{revtex4-1}
\usepackage[T1]{fontenc}
\usepackage[latin9]{inputenc}
\usepackage{color}
\usepackage{babel}
\usepackage{amsmath}
\usepackage{amssymb}
\usepackage{graphicx}
\usepackage[colorlinks=true,
    linkcolor=blue,
    citecolor=blue,
    urlcolor =blue]{hyperref}
\makeatletter
\usepackage{amsfonts}
\usepackage{babel}
\usepackage{braket}
\makeatother

\begin{document}

\title{Quantum Spin Dimers from Chiral Dissipation in Cold-Atom Chains}

\author{Tom\'as Ramos}
\email{tomas.ramos@uibk.ac.at}
\selectlanguage{english}
\affiliation{Institute for Quantum Optics and Quantum Information of the Austrian
Academy of Sciences, 6020 Innsbruck, Austria}
\affiliation{Institute for Theoretical Physics, University of Innsbruck, 6020 Innsbruck, Austria}

\author{Hannes Pichler}
\affiliation{Institute for Quantum Optics and Quantum Information of the Austrian
Academy of Sciences, 6020 Innsbruck, Austria}
\affiliation{Institute for Theoretical Physics, University of Innsbruck, 6020 Innsbruck, Austria}

\author{Andrew J. Daley}
\affiliation{Department of Physics and SUPA, University of Strathclyde, Glasgow
G4 0NG, United Kingdom}
\affiliation{Department of Physics and Astronomy, University of Pittsburgh, Pittsburgh, Pennsylvania 15260, USA}

\author{Peter Zoller}
\affiliation{Institute for Quantum Optics and Quantum Information of the Austrian
Academy of Sciences, 6020 Innsbruck, Austria}
\affiliation{Institute for Theoretical Physics, University of Innsbruck, 6020 Innsbruck, Austria}

\begin{abstract}
We consider the nonequilibrium dynamics of a driven dissipative spin
chain with chiral coupling to a one-dimensional (1D) bosonic bath, and its atomic implementation with a two-species mixture of cold quantum gases. The reservoir is represented by a spin-orbit coupled 1D quasicondensate of atoms in a magnetized phase, while the spins are identified with motional states of a separate species of atoms in an optical lattice. The chirality of reservoir excitations allows the spins to couple differently to left- and right-moving modes, which in our atomic setup can be tuned from bidirectional to purely unidirectional. Remarkably, this leads to a pure steady state in which pairs of neighboring spins form dimers that decouple from
the remainder of the chain. Our results also apply to current
experiments with two-level emitters coupled to photonic waveguides. 
\end{abstract}
\maketitle
In an open quantum many-body system, the competition of particle interactions,
external driving and the dissipative coupling to a quantum
reservoir can result in novel scenarios for the formation of strongly
correlated quantum states \cite{Muller:2012wh}. This is not only of interest as a nonequilibrium condensed matter problem \emph{per se} \cite{Diehl:2008aa,Scelle:2013in,Chen:2014it,Knap:2013el,Prosen:2010ju,Rao:2013de,Carr:2013hc,Honing:2013fx}, but dissipatively prepared entangled states also provide a potential resource for quantum information
tasks \cite{Stannigel:2012jk,Krauter:2011fj,Barreiro:2011jq,Lin:2013cc,Weimer:2010ez,Verstraete:2009kc}. 
Quantum optical systems of cold atoms or solid-state impurities provide a natural setting for such open many-body quantum systems.
The paradigmatic example is given by an ensemble of two-level atoms
driven by laser light, and coupled to a \emph{photonic} reservoir \cite{Chang:2012co,Reitz:2013bs,{Yalla:2014eb},Thompson:2013hx}, e.g., as one-dimensional (1D) engineered photonic band gap materials \cite{Goban:2013wp}. These model systems can be described as a collection of spin-$1/2$ systems, which
via the photonic modes interact with long-range dipole-dipole interactions, and exhibit collective and enhanced decay into radiation modes of photonic structures. The realization of such Dicke-type models \cite{Walls:1978hh,Gross:1982js}
coupled to low-dimensional quantum reservoirs, and the observation
of the associated dynamical quantum phases and phase transitions are, at present, an outstanding challenge in quantum optics \cite{GonzalezTudela:2013hn,{Baumann:2010js},{Zou:2014cy},{vanLoo:2013df}}.

\begin{figure}[b]
\includegraphics[width=0.472\textwidth]{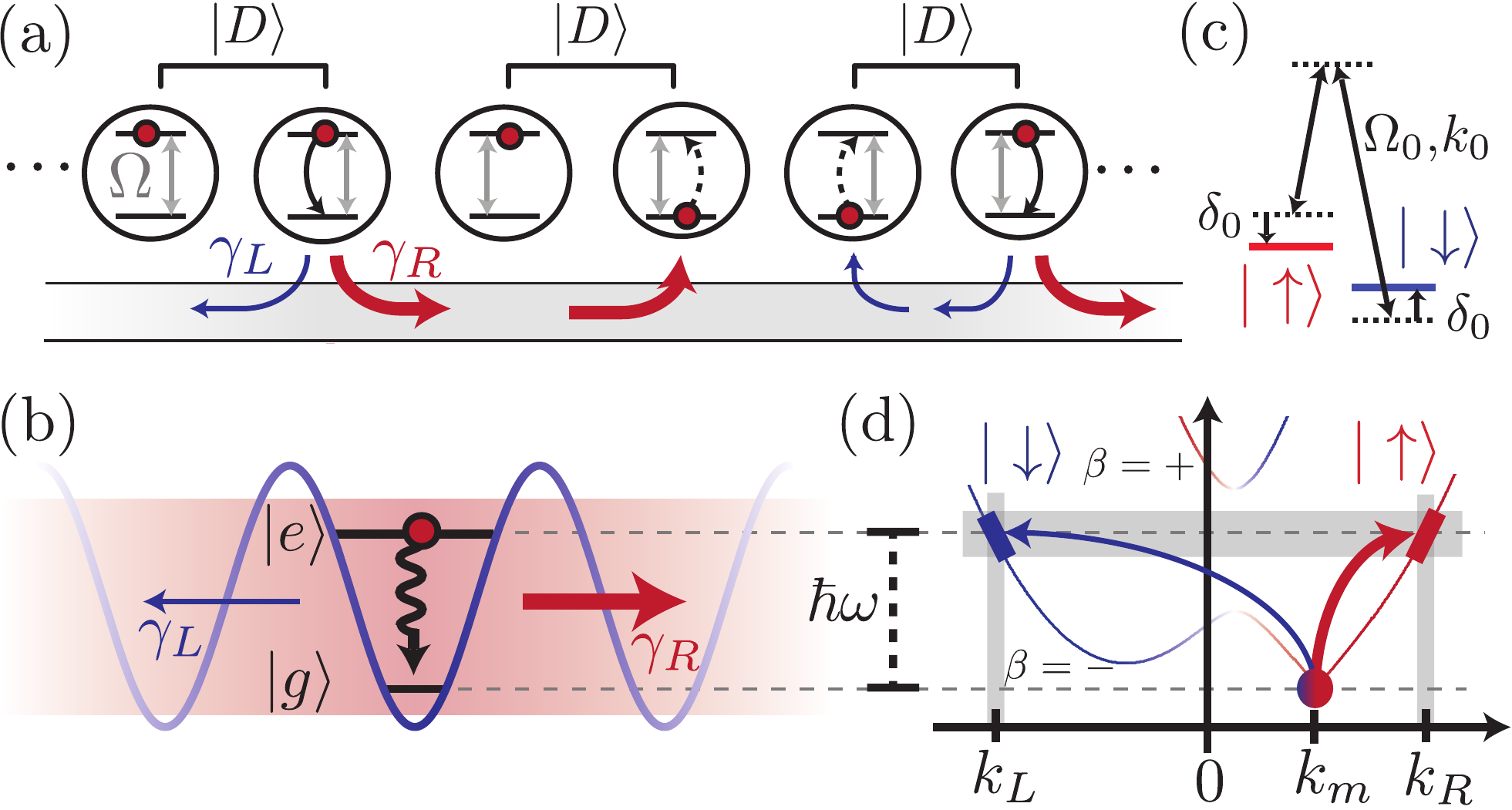}
\caption{The 1D spin chain coupled to a 1D chiral bosonic reservoir. (a) Driven spins decay into right- and left-moving reservoir modes with rates $\gamma_R$ and $\gamma_L$. For $\gamma_R\!\ne\!\gamma_L$, quantum spin dimers (indicated by $\ket{D}$) are formed as the unique pure steady state. (b-d) Implementation with a two-species mixture of cold atoms. (b) Spins are represented by the two lowest vibrational states of atoms $a$ on each site of a 1D optical lattice, which can ``decay'' due to collisions with a 1D SOC quasi-BEC, representing the bath. (c) SOC of atoms $b$ due to coupling of two internal states $\ket{\uparrow}$ and $\ket{\downarrow}$ via a Raman process \cite{Lin:2011hn}. (d) Dispersion relations $\hbar\omega_{k\beta}$ of the bath excitations in the plane wave phase. The red and blue arrows indicate excitations of atoms $b$ from the quasi-BEC (circle at $k_m$) to wave vectors $k_L$ and $k_R$, resonant with $\hbar\omega$.}
\label{fig:chain}
\end{figure}

In the present work, we introduce a realization of 
dissipative quantum magnetism based on cold atoms in optical lattices \cite{Daley:2004be,Schwager:2013ey}, where the quantum reservoir is represented by \emph{phononic} degrees of freedom of a 1D spin-orbit coupled Bose-Einstein quasicondensate (quasi-BEC) \cite{Lin:2011hn,Cheuk:2012id,Ji:2014jh,Wang:2012gv,Goldman:2013uq,Kruger:2010eg,Jacqmin:2011jp}. This model system provides a faithful and experimentally realistic representation of a chain of driven spin-$1/2$ particles coupled to a 1D bosonic bath. Crucially, spin-orbit coupling (SOC) makes the reservoir \emph{chiral}, with
the spins coupling differently to the left and right propagating
modes, $\gamma_L\!\ne\!\gamma_R$ [cf.~Fig.\,\ref{fig:chain}(a)]. This asymmetry is, moreover, tunable via the atomic parameters, making it possible to engineer the spin-bath coupling from purely unidirectional to fully bidirectional. 

To describe the dynamics of our 1D spin chain, we derive a quantum optical master equation for the reduced system density matrix $\rho(t)$, tracing over the reservoir degrees of freedom. This equation contains both long-range dipolar spin interactions, as mediated by the exchange of Bogoliubov excitations, and collective dissipative terms. Remarkably, at long times the system evolves to a pure many-body state of quantum spin dimers, $\rho(t)\xrightarrow{t\rightarrow\infty}|\Psi\rangle\langle\Psi|$
with $|\Psi\rangle\!=\!\bigotimes_{j=1}^{N/2}\ket{D}_{2j-1,2j}$. Here
\begin{equation}\label{eq:dimer}
\ket{D}_{j,l}\!\equiv\!\!\frac{1}{\sqrt{1+|\alpha|^2}}\!\!\left[\ket{g}_{j}\!\ket{g}_{l}\!+\!\!\frac{\alpha}{\sqrt{2}}\!\left(\!\ket{g}_{j}\!\ket{e}_{l}\!-\!\ket{e}_{j}\!\ket{g}_{l}\right)\!\right]\!\!
\end{equation}
is the spin-dimer state of a pair of spin-$1/2$ particles at lattice sites $j,l$ with $\ket{g},\ket{e}$ denoting the corresponding ground and excited states, and $\alpha$ a parameter defined below. This result is valid for a generic range of parameters in the case of reservoirs with broken left-right symmetry and an even number of spins [cf.\,Fig.\,\ref{fig:chain}(a)]. Further, it is also of immediate relevance in the context of recent proposals and experiments for two-level systems (TLSs) coupled to a photonic chiral reservoir \cite{Petersen:2014ki,Mitsch:2014tr,Sollner:2014wb,Young:2014ty}.

\emph{Model.---} We realize a driven dissipative spin chain coupled to a 1D bosonic reservoir with a two-species mixture of quantum gases. The corresponding setup is shown in Figs.\,\ref{fig:chain}(b)-\,\ref{fig:chain}(d). The spin chain is represented by spinless atoms of a first species $a$ (with mass $m_a$), trapped in a species-selective 1D optical lattice \cite{Scelle:2013in} of period $d$ [cf.~Fig.\,\ref{fig:chain}(b)]. We assume filling with one atom per site and a deep lattice to completely suppress the tunneling (Mott insulator). Thus, the ground and first vibrational states of the atom at lattice site $j$ with position $x_j$ represent a TLS, $\ket{g}_{j}$ and $\ket{e}_{j}$, or effective spin-$1/2$. Other vibrational states are decoupled due to the lattice anharmonicity. We can drive these TLSs near their transition frequency $\omega$ via a Raman process with frequency $\nu$ and Rabi frequencies $\Omega_{j}$. In the rotating wave approximation (RWA), the Hamiltonian for the driven spin chain with $N$ atoms reads ($\sigma_{j}\equiv\ket{g}_{j}\bra{e}$)
\begin{align}
H_{{\rm sys}}=\hbar\omega\sum_{j=1}^{N}\sigma_{j}^{\dag}\sigma_{j}+\hbar\sum_{j=1}^{N}\left(\Omega_{j}\sigma_{j}e^{i\nu t}+{\rm H.c.}\right).\label{Hsys}
\end{align}

The 1D bosonic quantum reservoir is realized with a second atomic species $b$ (with mass $m_b$). We assume, again, trapping in a 1D geometry (aligned with the optical lattice), however, with the atoms $b$ now moving freely along a homogeneous 1D wire. In addition, we prepare them in the quasi-BEC regime \cite{Petrov:2000cf,Andersen:2002cn,Mora:2003kb,Kruger:2010eg,Jacqmin:2011jp}; i.e., the linear density $\bar{\rho}$ satisfies $\hbar^2\bar{\rho}^2/m_b\gg k_BT,\mu$, with $T$ the temperature and $\mu$ the chemical potential. Atoms $a$ will couple to the reservoir atoms $b$ via collisional interactions. In particular, there will be resonant processes, where an atom $a$ ``decays'' from $\ket{e}$ to $\ket{g}$, creating an excitation of energy $\hbar\omega$ in the reservoir gas \cite{{Daley:2004be},Griessner:2006hg,Griessner:2007hf} [cf.~Fig.\,\ref{fig:chain}(b)]. These excitations will propagate along the wire and represent the right- and left-moving bosonic excitations constituting our 1D bath. First experiments along these lines have been realized with a three-dimensional BEC as the reservoir \cite{Scelle:2013in,Chen:2014it}.

\begin{figure}[t]
\includegraphics[width=0.48\textwidth]{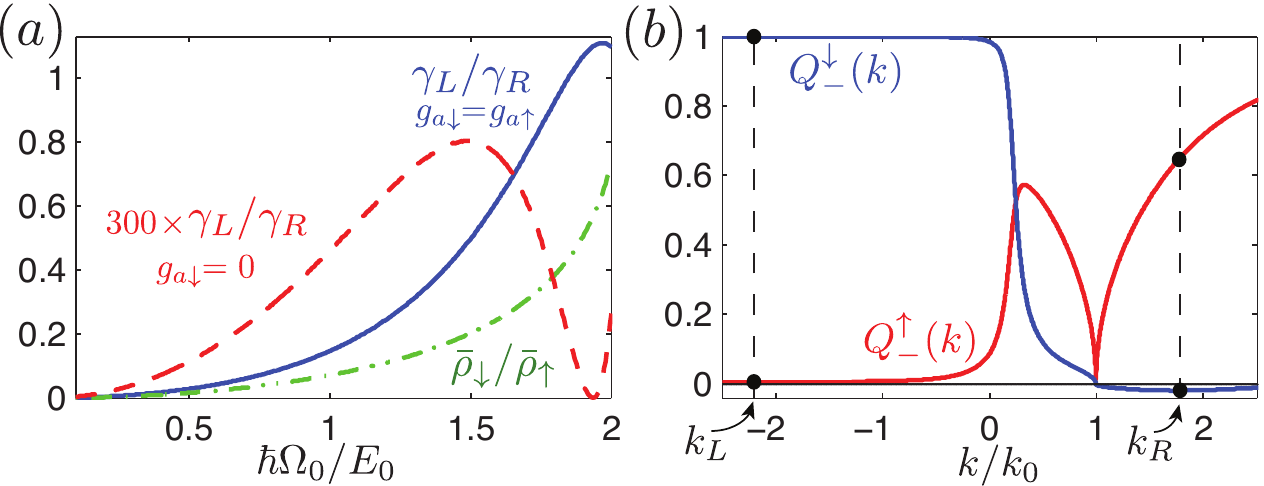}
\caption{Tunability of decay asymmetry into chiral left- and right-moving modes (a) $\gamma_L/\gamma_R$ as a function of $\hbar\Omega_0/E_0$ for $g_{a\downarrow}\!\!=\!\!g_{a\uparrow}$ (solid line) and $g_{a\downarrow}\!\!=\!\!0$ (dashed line). The dash-dotted line shows the reservoir spin polarization $\bar{\rho}_{\downarrow}/\bar{\rho}_{\uparrow}$. (b) Density fluctuation coefficients $Q^{\lambda}_{-}(k)$ ($\lambda\!\!=\uparrow,\downarrow$) in the lower branch for $\hbar\Omega_0\!=\!0.2E_0$. The wave vectors for left- and right-moving excitations $k_s$ ($s\!\!=\!\!L,R$) are indicated, where $Q^{\lambda}_{-}(k_s)$ show their strong spin polarization. Other parameters are $\bar{\rho}\!=\!6.14k_0$, $m_a/m_b\!=\!2$, $g_{\uparrow\uparrow}\!=\!g_{\uparrow\downarrow}\!=\!g_{\downarrow\downarrow}\!=\!0.23E_0/k_0$, $g_{a\uparrow}\!=\!-0.37E_0/k_0$, $\hbar\omega\!=\!1.46E_0$ and $\delta_0\!=\!-0.004E_0$.}
\label{fig:asymmetry}
\end{figure}

A chiral reservoir with asymmetric decay of spins to left- and right-moving modes ($\gamma_L\!\ne\!\gamma_R$) is obtained by adding SOC to the 1D quasi-BEC. Following Ref.~\cite{Lin:2011hn}, SOC with equal Rashba and Dresselhaus contributions can be implemented by coupling two internal states $\ket{\uparrow}$ and $\ket{\downarrow}$ of the reservoir atoms $b$ via Raman lasers with momentum transfer $2\hbar k_0$, coupling strength $\Omega_{0}$, detuning $2\delta_{0}$ and recoil energy $E_0\equiv \hbar^2k_0^2/(2m_b)$ [cf.~Fig.\,\ref{fig:chain}(c)]. Using an extension of Bogoliubov theory to quasicondensates \cite{Mora:2003kb,Petrov:2000cf}, one can diagonalize the reservoir Hamiltonian in terms of Bogoliubov-like excitations as $H_{{\rm res}}\!=\!\sum_{k,\beta}\hbar\omega_{k\beta}b_{k\beta}^{\dag}b_{k\beta}$. We refer to Supplemental Material in Ref.~\cite{SUPPL} for details. Here $b_{k\beta}$ are bosonic annihilation operators for excitations with wave vector $k$ in the branch $\beta\!=\!\pm$, and $\omega_{k\beta}$ is the corresponding excitation spectrum shown in Fig.\,\ref{fig:chain}(d) for $\hbar\Omega_0\!\ll\!E_0$. What is crucial for our proposal is that at energies $\sim E_0$, there is an energy window $\sim\!\Omega_0$ in which excitations are chiral; i.e., all excitations with positive group velocity are strongly polarized along $\ket{\uparrow}$, while the ones with negative group velocity are strongly polarized along $\ket{\downarrow}$. This locking of the propagation direction to the spin is reminiscent of chiral edge modes in systems with artificial gauge fields \cite{Atala:2014gf,Celi:2014dg}. To be specific, the excitation spectrum of Fig.\,\ref{fig:chain}(d) is obtained when the SOC quasi-BEC is prepared in the so-called plane wave phase \cite{{Ji:2014tr},{Li:2012dd},{Martone:2012kl}} with quasicondensation at a positive wave vector $k_m$. This can be achieved by using a finite detuning $\delta_0<0$, satisfying $\bar{\rho}(g_{\uparrow\uparrow}-g_{\uparrow\downarrow})/2<\hbar|\delta_{0}|\ll E_0$, where $g_{\uparrow\uparrow}, g_{\downarrow\downarrow}, g_{\uparrow\downarrow}\!\geq\!0$ are the 1D collisional interaction parameters of the reservoir gas (cf.~\cite{SUPPL}). An important characteristic of this phase is that the atoms in the quasi-BEC are spin polarized, as manifested by  $\bar{\rho}_{\downarrow}/\bar{\rho}_{\uparrow}<1$, where $\bar{\rho}_{\uparrow}$ and $\bar{\rho}_{\downarrow}$ are the mean densities of the different quasi-BEC spin components ($\bar{\rho}=\bar{\rho}_{\uparrow}+\bar{\rho}_{\downarrow}$). A feature of the synthetic SOC is the tunability of this spin polarization with $\Omega_0$ [cf.~Fig.\,\ref{fig:asymmetry}(a)].

We take a quantum optical point of view in describing the system-bath interaction, which is motivated by the analogy with TLSs coupled to a 1D photonic bath in the weak coupling limit. Microscopically, it is given in our setup by collisional interactions between $a$ and $b$ atoms. For spinless atoms $a$, these collisions are spin conserving and reduce to interspecies density-density interactions. Therefore, density fluctuations of the reservoir atoms in a frequency band around $\omega$ provide an energy-conserving mechanism for spin decay. In terms of elementary excitations, they can be written as $\delta\rho_{\lambda}\!=\!\sqrt{\bar{\rho}_{\lambda}/L}\sum_{k,\beta=\pm}Q_{\beta}^{\lambda}(k)b_{k\beta}e^{i(k-k_m)x}+{\rm H.c.}$ \cite{Mora:2003kb,SUPPL}, where $L$ is a quantization length and the coefficients $Q_{\beta}^{\lambda}(k)$ (with $\lambda\in\{\uparrow,\downarrow\}$) reflect the spin-polarization of excitations [cf.~Fig.\,\ref{fig:asymmetry}(b)]. By placing the TLS transition frequency $\omega$ in the aforementioned energy window around $E_0$, the RWA restricts the reservoir to chiral excitations only, provided $\gamma_L,\gamma_R\ll \Omega_0,\omega$. Furthermore, we can linearize the dispersion in intervals $I_L$ and $I_R$, around the corresponding resonant wave vectors $k_{L}$ and $k_{R}$, with group velocities $v_{L}\!<\!0$ and $v_{R}\!>\!0$ [cf.~Fig.\,\ref{fig:chain}(d)]. As a result, the interaction Hamiltonian can be written in a form reminiscent of the prototypical quantum optical RWA Hamiltonian as (cf.~Ref.~\cite{SUPPL})
\begin{align}
H_{{\rm int}}=&i\hbar\!\!\!\sum_{s=L,R}\!\sqrt{\frac{\gamma_{s}|v_{s}|}{L}}\sum_{k\in I_s,j}\!\sigma_{j}^{\dag}b_{k,-}e^{i(k-k_m)x_{j}}\!+\!{\rm H.c.},\label{Hint}
\end{align}
with decay rates into the left and right propagating modes ($s\!=\!R,L$) given by 
\begin{align}
\gamma_{s}\equiv\frac{\eta(k_s)e^{-\eta(k_s)}}{\hbar^{2}|v_{s}|}\bigg(\sum_{\lambda=\uparrow,\downarrow}\!g_{a\lambda}\sqrt{\bar{\rho}_\lambda}
Q_{-}^{\lambda}(k_s) \bigg)^2.\label{decays}
\end{align}
Here $g_{a\uparrow},g_{a\downarrow}$ are the collisional couplings between $a$ and $b$ atoms, and $\eta(k)\equiv(E_0/\hbar\omega)(m_b/m_a)[(k-k_m)/k_0]^2$.

The physical origin of the decay asymmetry $\gamma_R\ne \gamma_L$ is primarily the preparation of the reservoir in the plane wave phase at $k_m\!>\!0$. For $\hbar\Omega_0\!\ll\! E_0$, the reservoir atoms are strongly spin polarized $\bar{\rho}_{\uparrow}\!\gg\!\bar{\rho}_{\downarrow}$ [cf.~Fig.\,\ref{fig:asymmetry}(a)], suppressing the creation of left-moving excitations in the spin-conserving collisions due to the small overlap of the spin wave functions. In addition, creating left- or right-moving excitations requires different momentum transfers [cf.~Fig.\,\ref{fig:chain}(d)] which also give rise to an asymmetry, reflected by the coupling constants $\eta(k_s)$. As illustrated in Fig.\,\ref{fig:asymmetry}(a), the decay asymmetry can be tuned with $\Omega_0$ from essentially unidirectional $\gamma_L/\gamma_R\ll 1$ to fully bidirectional $\gamma_L/\gamma_R=1$. Another mechanism for an asymmetry is provided in the case of spin-dependent collisions ($g_{a\uparrow}\neq g_{a\downarrow}$). In particular, for $g_{a\uparrow}\gg g_{a\downarrow}$, there is predominant decay to the right-moving modes. Remarkably, there are parameters for which $Q_{-}^{\uparrow}(k_L)\!=\!\!0$ (cf.~Ref.~\cite{SUPPL}), making it possible to realize an ideal cascaded spin chain with $\gamma_L=0$, if $g_{a\downarrow}\!=\!0$ [cf.~Fig.\,\ref{fig:asymmetry}(a)].

\emph{Master equation.---} We derive a master equation for the reduced density operator $\rho(t)$ of the spin chain by eliminating the reservoir atoms in the Born-Markov approximation \cite{QuantumNoise,{Daley:2004be}}. For $\hbar\omega\!\gg\!k_BT$, and neglecting retardation effects provided $\gamma_s\!\ll\!2\pi|v_s|/(Nd)$ \cite{Chang:2012co,Milonni:1974bo}, we find
\begin{align}
\dot{\rho}&=-(i/\hbar)[H_{{\rm sys}},\rho]+\mathcal{L}_{\rm B}\rho+\mathcal{L}_{\rm C}\rho,\label{ME}
\end{align}
where $H_{\rm sys}$ is defined in Eq.\,(\ref{Hsys}) and the Liouvillian terms describing reservoir-mediated interactions read
\begin{align*}
&\mathcal{L}_{\rm B}\rho\!\equiv\!\gamma_L\!\sum_{j,l}\!\left[-i\sin(|\phi_{jl}|)[\sigma_{l}^{\dag}{\sigma}_{j},\rho]\!+\!\cos(|\phi_{jl}|){\cal D}({\sigma}_{j},{\sigma}_{l})\rho\right]\!,\\
&\mathcal{L}_{\rm C}\rho\!\equiv\!\frac{\Delta\gamma}{2}\!\sum_{j}\!{\cal D}({\sigma}_{j},{\sigma}_{j})\rho+\!\Delta\gamma\sum_{j>l}\!\left(\!e^{-i\phi_{jl}}[{\sigma}_{j},\rho{\sigma}_{l}^{\dag}]\!+\textrm{H.c.}\right)\!.
\end{align*}
In writing Eq.\,(\ref{ME}) we used the notation ${\cal D}(a,b)\rho\equiv 2a\rho b^{\dag}-b^{\dag}a\rho-\rho b^{\dag}a$ and assumed $\Delta\gamma\equiv\gamma_R-\gamma_L\geq 0$. Additionally, we defined phase factors $\phi_{jl}\equiv(x_j-x_l)(k_{R}-k_{L})/2$, and redefined $\sigma_j\rightarrow\sigma_je^{-i(k_{R}+k_L-2k_m)x_j/2}$ and $\Omega_j\rightarrow\Omega_je^{i(k_{R}+k_L-2k_m)x_j/2}$. The Liouvillian $\mathcal{L}_B$ is familiar from TLSs coupled to a symmetric (bidirectional) 1D waveguide \cite{Chang:2012co,{GonzalezTudela:2013hn}}. It contains a coherent (Hamiltonian) part, describing infinite-range dipole-dipole interactions and an incoherent part with ``quantum jump operators'' \cite{QuantumNoise} associated with infinite-range superradiant collective decay. Its strength is given by the smaller of the decay rates $\gamma_L$. The last term, $\mathcal{L}_C$, is the Liouvillian of a {\em cascaded quantum system} \cite{QuantumNoise,Stannigel:2012jk}, i.e., where bath excitations can only move to the right. Its strength is given by $\Delta\gamma$ and thus it appears only if the left-right symmetry is broken.

\emph{Quantum spin dimers as the steady state.---} We consider a situation where the lattice spacing $d$ is commensurate with the wavelength of the reservoir excitations, $(k_R-k_L)d= 4\pi n$ ($n$ is an integer \cite{footnote}), so that the dipole-dipole interactions vanish. In addition, we assume that all spins are driven homogeneously, $\Omega_j\!=\!\Omega$, and on-resonance, $\nu\!=\!\omega$.

We note that for $\Delta\gamma\!=\!0$, Eq.\,(\ref{ME}) reduces to a totally symmetric Dicke model, where a nonequilibrium quantum phase transition at a critical driving $\Omega_c\!\equiv\!N\gamma_L/4$ has been predicted \cite{Walls:1978hh,GonzalezTudela:2013hn}. In this case, only coupling within the so-called Dicke manifolds is allowed, which leads to multiple steady states. In contrast, when $\Delta\gamma\neq 0$ this symmetry is broken and the steady state is unique. Remarkably, for an even number of spins, the steady state is pure and it dimerizes; i.e., each spin pairs up with one of its neighbors in the entangled state $\ket{D}$ given in Eq.\,\eqref{eq:dimer} with the singlet fraction $\alpha= 2i\sqrt{2}\Omega^{\ast}/\Delta\gamma$. Such a dimerized state represents  a dark state of the driven-dissipative many-body dynamics \cite{{Parkins:1993ko}}, where excitations are exchanged between two adjacent spins, but they do not escape from the pair due to quantum interference. For the ideal cascaded case ($\gamma_L=0$), Ref.\,\cite{Stannigel:2012jk} has previously discussed such ``cooling to dimers'' with engineered optomechanical systems. In Ref.~\cite{SUPPL}, we give a formal proof that this dimerization is in fact the generic steady state of Eq.\,(\ref{ME}) for the whole range $0\leq\gamma_L/\gamma_R<1$.

\begin{figure}[t]
\includegraphics[width=0.48\textwidth]{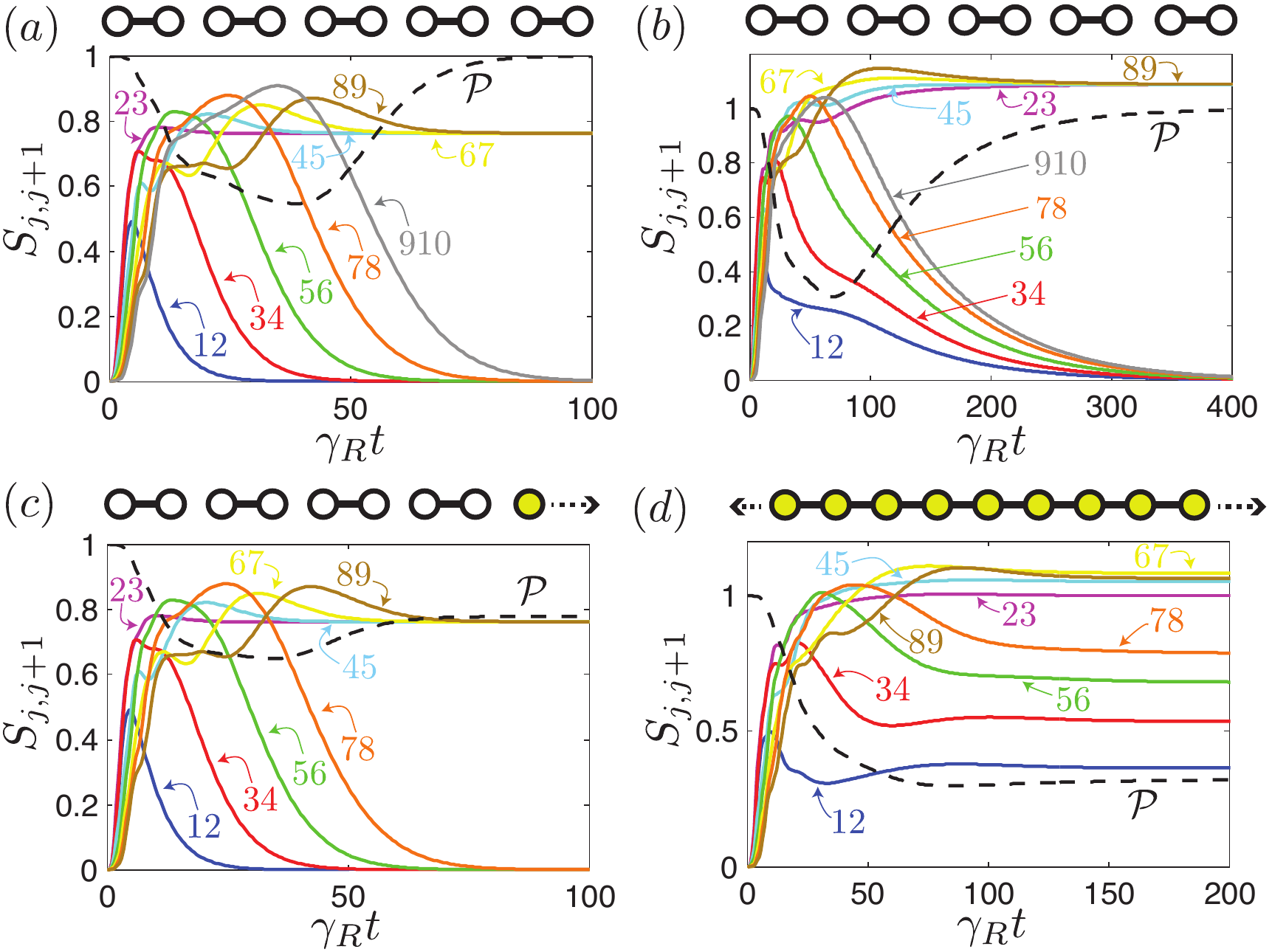}
\caption{Dynamical formation of spin dimers as the unique steady state of the driven-dissipative spin chain. We plot the entropy $S_{j,j+1}(t)$ of all adjacent spin pairs (solid lines) and the purity ${\cal P}(t)$ of the total state (black dashed line) for the initial condition $|\Psi(0)\rangle\!=\!\bigotimes_{j=1}^{N}\ket{g}_j$. Results are shown for $\Omega=0.5\gamma_R$ and (a) $N\!=\!10$, $\gamma_L\!=\!0$, (b) $N\!=\!10$, $\gamma_L\!=\!0.4\gamma_R$, (c) $N\!=\!9$, $\gamma_L\!=\!0$, (d) $N\!=\!9$, $\gamma_L\!=\!0.4\gamma_R$.}
\label{fig:time_evolution}
\end{figure}

To gain insight into how a spin chain dynamically purifies and arranges itself into dimers, we numerically calculate the time evolution of the purity of the total state ${\cal P}\!\equiv\!{\rm Tr}\lbrace\rho^2\rbrace$, and the entropy of adjacent spin pairs $S_{j,j+1}\!\equiv\!-{\rm Tr}\lbrace\rho_{j,j+1}\ln(\rho_{j,j+1})\rbrace$. Here $\rho_{j,l}$ is the reduced density operator for spins $j$ and $l$. The formation of pure dimers is manifested by ${\cal P}(t)\!\rightarrow\!1$ and $S_{2j-1,2j}(t)\rightarrow\!0$, $\forall j=1,...,N/2$, as shown in Figs.\,\ref{fig:time_evolution}(a) and \ref{fig:time_evolution}(b). For any ratio $\gamma_L/\gamma_R<1$, pairs are purified ``from left to right,'' but only in the cascaded limit does this happen successively at a constant speed [cf.~Fig.\,\ref{fig:time_evolution}(a)]. The time scale $t_{\rm ss}$ to reach the steady state increases with $\gamma_L/\gamma_R$. In the limit $\gamma_L/\gamma_R\!\rightarrow\!1$, we numerically find the scaling $t_{\rm ss}\!\sim \!(1-\gamma_L/\gamma_R)^{-4}$ for small system sizes (cf.~Ref.\cite{SUPPL}). 
When the number of spins is odd, it is not possible for all of them to pair up in dimers. Nevertheless, in the cascaded limit dimers are still formed, leaving only the last unpaired spin in a mixed state [cf.~Fig.\,\ref{fig:time_evolution}(c)]. The excitations emitted by this last spin propagate only to the right and do not affect the dimers on its left. On the other hand, if excitations can also propagate to the left, no dimers are formed because the output of an unpaired spin breaks them up [cf.~Fig.\,\ref{fig:time_evolution}(d)].
   
\begin{figure}[t]
\includegraphics[width=0.48\textwidth]{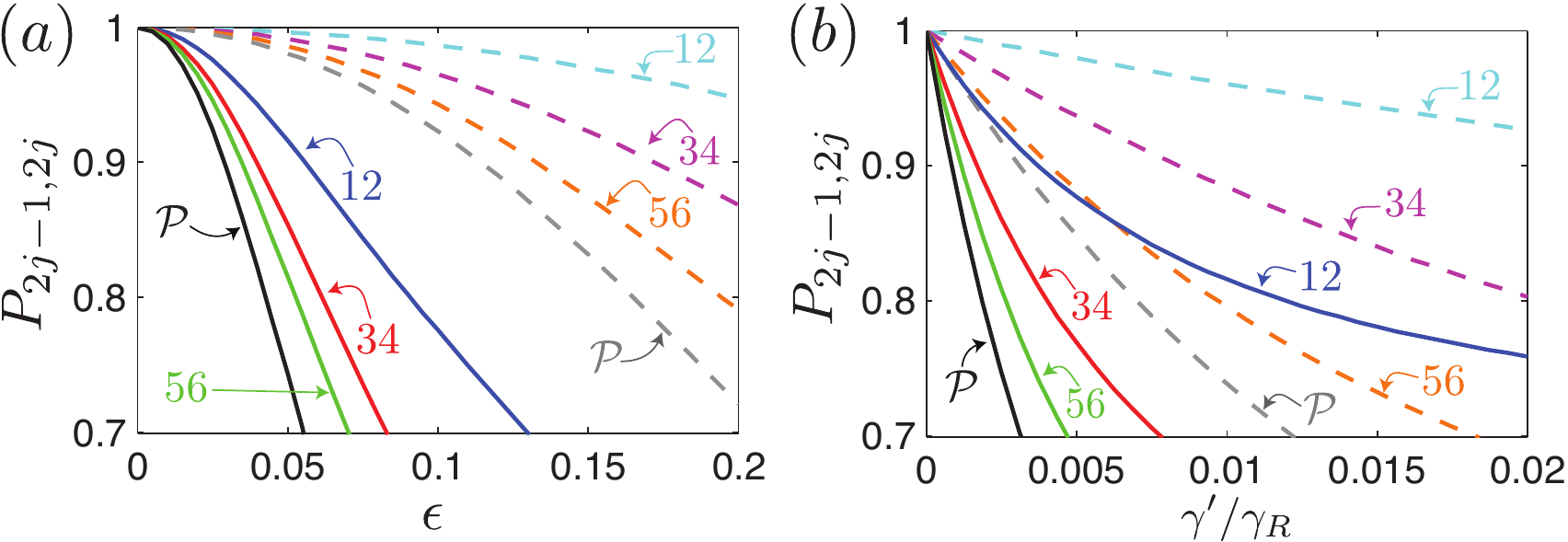}
\caption{Robustness of the dimerized steady state against imperfections for $N\!=\!6$. (a) Pair purities ${\cal P}_{2j-1,2j}$ and total purity ${\cal P}$ as a function of $\epsilon$ (see text), for $\gamma_L\!=\!0.1$ (dashed line) and $\gamma_L\!=\!0.4\gamma_R$ (solid line). (b) ${\cal P}_{2j-1,2j}$ and ${\cal P}$ as a function of decay outside the 1D bath $\gamma'$, for $\gamma_L\!=\!0$ (dashed line) and $\gamma_L\!=\!0.4\gamma_R$ (solid line). We fix $\Omega=\!0.5\gamma_R$.
}\label{fig:imperfections}
\end{figure}

To ensure robustness of dissipative dimerization, we studied numerically the effect of various imperfections on the steady state of Eq.\,\eqref{ME}, reflected by the pair purities ${\cal P}_{2j-1,2j}\!\equiv\!{\rm Tr}\{(\rho_{2j-1,2j})^2\}$. In general, imperfections give rise to an incomplete decoupling of spin pairs from the rest of the chain, as Fig.\,\ref{fig:imperfections}(a) illustrates for deviations from the commensurability condition, quantified by $\epsilon\equiv(k_R-k_L)d-4\pi n$. We observe particular robustness for low $\gamma_L/\gamma_R$ and a decrease in the pair-purities from left to right. However, already for $\epsilon\lesssim 0.1$ we obtain ${\cal P}_{2j-1,2j}\gtrsim 0.9$, when $\gamma_L=0.1\gamma_R$. Qualitatively, the same behavior is observed for deviations in the detuning and phases of the coherent driving field. On the other hand, on-site decay outside the 1D reservoir leads to a significant decrease of the purities [cf.~Fig.\ref{fig:imperfections}(b)]. This could be a concern for implementations with photonic waveguides \cite{Reitz:2013bs,Goban:2013wp}. However, in the setup proposed here, such processes are only weakly induced (e.g., by classical noise \cite{Pichler:2013fw}), and thus they expected to be negligible compared with $\gamma_R$.

\emph{Estimates.---} We consider a quasi-BEC of ${}^{87}$Rb with $E_0/(2\pi\hbar)\!\approx\!3.5\,{\rm kHz}$, $2|\delta_0|\!\gtrsim\!25\,{\rm Hz}$ \cite{{Lin:2011hn}}, $T\!=\!5\,nK$, $\bar{\rho}\!=\!48\,\mu{\rm m}^{-1}$ (e.g.~with 4800 atoms confined to $L\!\sim\!100\,\mu{\rm m}$ \cite{{Jacqmin:2011jp},{Kruger:2010eg},{Meyrath:2005dl}}), and a transverse trapping frequency $\omega_{\perp}/(2\pi)\!=\!10\,{\rm kHz}$. For the spin chain we consider Yb, because it is spinless and heavy ($m_a/m_b\approx 2$) (cf.~Ref.~\cite{SUPPL}). With interspecies scattering lengths between ${}^{87}$Rb and ${}^{172}$Yb of $a_{a\uparrow}\!=\!a_{a\downarrow}\!\approx\!-160.7\,a_{\rm Bohr}$ \cite{Borkowski:2013kr} and $\hbar\omega/(2\pi)\approx 5.3\,{\rm kHz}$, one obtains decay rates $\gamma_R/(2\pi)\!\sim\!100\,{\rm Hz}$, with asymmetries $10^{-3}\!<\!\gamma_L/\gamma_R\!<\!1.1$ [cf.~Fig.\,\ref{fig:asymmetry}(a)]. These rates validate \emph{a posteriori} the RWA and Markov approximations, as well as neglecting retardation effects for systems up to $N\!\sim\! 30$ spins spaced by $d\!\sim 800\,{\rm nm}$. On the other hand, these ``quantum optical'' approximations can also be deliberately violated in our setup to study retardation and non-Markovian effects outside the validity of the master equation treatment. We note that heating due to photon scattering \cite{Goldman:2013uq} in the $^{87}$Rb SOC quasi-BEC is negligible on time scales related to the formation of dimers (cf.~\cite{SUPPL,heating}).

\emph{Outlook.---} We have shown how SOC in an atomic gas can be used to engineer a chiral reservoir for spin chains. The tunable asymmetry of the coupling to left- and right-moving excitations leads to a pure steady state in which neighboring spins are dimerized, representing a novel form of \textit{dissipative} quantum magnetism \cite{valencebond,Auerbach:1994jf}. While the cold-atom realization provides particular advantages, our results also apply to implementations with photons \cite{Mitsch:2014tr,Sollner:2014wb,Young:2014ty}. We have shown \cite{Pichler:2014ab} that the present results generalize to the dissipative formation of pure many-body states of spin-$1/2$ tetramers, hexamers, etc., by appropriate driving patterns \cite{Stannigel:2012jk}. This multipartite entanglement can be detected via the Fisher information \cite{Hyllus:2012ha}, which recently has been measured in cold-atom experiments \cite{Strobel:2014eg}.

We thank M. A. Baranov, Y. Castin, N. Goldman, A. González-Tudela, Y. Li, G. I. Martone, C. Mora, J. V. Porto, S. L. Rolston, and K. Stannigel for helpful discussions. Work in Innsbruck was supported by the ERC Synergy Grant UQUAM, the EU grant SIQS and the Austrian Science Fund through SFB FOQUS. T.~R. further acknowledges financial support from BECAS CHILE.

\newpage
\onecolumngrid
\newpage
{
\center \bf \large 
Supplemental Material for: \\
Quantum Spin-Dimers from Chiral Dissipation in Cold Atom Chains\vspace*{0.1cm}\\ 
\vspace*{0.0cm}
}
\begin{center}
Tom\'as Ramos$^{1,2}$, Hannes Pichler$^{1,2}$, Andrew J. Daley$^{3,4}$, and Peter Zoller$^{1,2}$\\
\vspace*{0.15cm}
\small{\textit{$^1$Institute for Quantum Optics and Quantum Information of the Austrian
Academy of Sciences, 6020 Innsbruck, Austria\\
$^2$Institute for Theoretical Physics, University of Innsbruck, 6020 Innsbruck, Austria\\
$^3$Department of Physics and SUPA, University of Strathclyde, Glasgow
G4 0NG, UK\\
$^4$Department of Physics and Astronomy, University of Pittsburgh, Pittsburgh PA 15260, USA}}\\
\vspace*{0.25cm}
\end{center}

\twocolumngrid

\section{Diagonalization of the spin-orbit coupled reservoir Hamiltonian in the 1D quasi-condensate regime}

In this section we give an explicit expression for the effective 1D many-body reservoir Hamiltonian $H_{\rm res}$ of the main text, taking into account spin-orbit coupling (SOC) and contact interactions. Assuming the 1D quasi-BEC regime \cite{popovSM,Petrov:2000cfSM,Mora:2003kbSM,Andersen:2002cnSM}, we diagonalize $H_{\rm res}$ in terms of elementary (Bogoliubov-like) excitations by using an extension of Bogoliubov theory developed by Mora and Castin in Ref.\,\cite{Mora:2003kbSM}. In passing, we derive the expression for the density fluctuations in terms of Bogoliubov-like excitations also given in the main text.

\subsection{Many-body Hamiltonian for reservoir atoms}

The reservoir is composed of a gas of $M$ cold bosonic atoms with two internal states $\lbrace\ket{\uparrow}$, $\ket{\downarrow}\rbrace$, which are coupled via Raman lasers to realize artificial SOC with equal Rashba and Dresselhaus contributions \cite{Lin:2011hnSM}. In addition, we strongly confine the atoms in two directions such that the dynamics is effectively restricted to 1D \cite{Olshanii:1998jrSM} and we neglect the trapping potential in this remaining direction. Importantly, these reservoir atoms are not affected by the optical lattice potential also present in the setup [See Fig.\,1(b) of main the text], which can be realized by using a species-specific optical lattice \cite{Scelle:2013inSM}. As a result, the SOC reservoir atoms are freely moving along a 1D wire. Taking into account the contact interactions, the many-body Hamiltonian $H_{\rm res}$ reads
\begin{align}
H_{\rm res}&=\!\sum_{\nu,\lambda=\uparrow,\downarrow}\int\! dx\left[\psi^{\dag}_{\nu}H^{\rm SOC}_{\nu\lambda}\psi_{\lambda}+\frac{g_{\nu\lambda}}{2}\psi^{\dagger}_{\nu}\psi^{\dagger}_{\lambda}\psi_{\lambda}\psi_{\nu}\right],
\label{Hres}
\end{align}
where the field operators $\psi_{\nu}(x)$ with $\nu\!=\!\lbrace\uparrow,\downarrow\rbrace$ satisfy bosonic commutation relations $[\psi_{\nu}(x),\psi_{\lambda}^{\dag}(x')]=\delta_{\nu\lambda}\delta(x-x')$ and $g_{\nu\lambda}$ are the 1D-renormalized s-wave interaction parameters \cite{Olshanii:1998jrSM} between atoms in different internal states. In addition, $H^{\rm SOC}_{\nu\lambda}$ denote the components of the SOC Hamiltonian given by
\begin{align}
H^{\rm SOC}\equiv\frac{1}{2m_b}\!\left(\!-i\hbar\frac{\partial}{\partial x}-\hbar k_0\tau_z\right)^2\!\!\!+\hbar\Omega_0\tau_x+\hbar\delta_0\tau_z.\label{HSOC}
\end{align}
Here the $\tau$ symbols are the standard $2\times 2$ Pauli matrices and $m_b$ is the mass of the reservoir atoms. In current experiments \cite{Lin:2011hnSM}, this Hamiltonian (\ref{HSOC}) is implemented by coupling two hyperfine states (e.g.~$^{87}$Rb) via a Raman process with momentum transfer $2\hbar k_0$, coupling strength $\Omega_0$ and two photon detuning $2\delta_0$ and recoil energy $E_0\equiv \hbar^2k_0^2/(2m_b)$. Note that the time-independent form of the SOC Hamiltonian in Eq.\,(\ref{HSOC}) is not given in the lab frame, but rather in a spin-rotated frame after applying the unitary ${\cal U}\equiv e^{i(k_0x+\Delta\omega_0 t/2)\tau_z}$, with $\Delta\omega_0$ being the frequency difference between Raman lasers \cite{Martone:2012klSM}.

\subsection{Coarse-graining in position space}

For our later analysis of the interaction between lattice atoms and reservoir, it will be very useful to have a diagonal expression of the reservoir Hamiltonian (\ref{Hres}) in terms of elementary excitations on top of a macroscopic equilibrium configuration. Due to the large phase fluctuations in the 1D gas, there is no single macroscopically occupied state even at zero temperature \cite{Hohenberg:1967brSM,Kane:1967etSM} and thus the standard Bogoliubov treatment is not valid. Nevertheless, in the so-called quasi-condensate regime of weak interactions and very low temperature, the relative density fluctuations around a mean density are small, which allows for a systematic expansion and subsequent diagonalization of the many-body Hamiltonian $H_{\rm res}$ in a Bogoliubov-like manner. Several methods have been proposed in the literature to tackle this problem \cite{popovSM,Petrov:2000cfSM,Mora:2003kbSM,Andersen:2002cnSM}, but here we use the one developed by Mora and Castin in Refs.\,\cite{Mora:2003kbSM,Castin2004SM}, since it allows us to treat particle-like excitations with energies on the order of or greater than the chemical potential $\mu$. This approach has been mainly used to describe single-component low-dimensional Bose gases, but here we apply the same principles to a two component case with additional SOC, formally similar to the situation considered in Ref.\,\cite{Whitlock:2003ixSM}. We start by writing the field operators in a density-phase representation, $\psi_{\nu}(x)\equiv e^{i\theta_{\nu}}\sqrt{\rho_{\nu}}$, where $\rho_{\nu}(x)$ and $\theta_{\nu}(x)$ are the density and phase operators for each spin component $\nu=\lbrace\uparrow,\downarrow\rbrace$. It will be convenient to decompose the latter operators further as
\begin{align}
\rho_{\nu}=\bar{\rho}_{\nu}+\delta\rho_{\nu},\qquad \theta_{\nu}=\bar{\theta}_{\nu}+\delta\theta_{\nu},
\end{align}
where $\delta\rho_{\nu}(x)$ and $\delta\theta_{\nu}(x)$ describe the density and phase fluctuations of each spin component $\nu$ around the mean density and phase $\bar{\rho}_{\nu}(x)$ and $\bar{\theta}_{\nu}(x)$, respectively. In order to consistently define hermitian density and phase operators that approximately satisfy the standard commutation relations $[\rho_{\nu}(x),\theta_{\lambda}(x')]\approx i\delta_{\nu\lambda}\delta(x-x')$ and do not lead to divergences in the theory, Mora and Castin propose to apply a coarse-grained approximation in position space, assuming the large mean density limit. Additionally, for the direct application of the method in Ref.\,\cite{Mora:2003kbSM} to this two component case, here we also require a non-vanishing mean density for each spin separately $\bar{\rho}_{\nu}(x)\neq 0$, a condition that can always be met by a suitable change of spin basis in Eq.\,({\ref{Hres}}) [cf. Sec.\,\ref{rotatedbasis} for more details]. The procedure consists of discretizing the 1D space of length $L$ (with periodic boundary conditions) in small boxes of length $l$ for which the centers are located on a uniform grid at discrete positions $x$. The length $l$ must be chosen large enough such that there is a large mean number of particles in each box, but at the same time $l$ should be much smaller than all other relevant length scales of the system, so that the inclusion of the grid does not modify the physics of the continuous model. Therefore, the necessary inequalities read,   
\begin{align}
\bar{\rho}^{-1}\ll l\ll\ \xi,\lambda_T,\pi/k_{\rm max},\label{quasicondlength}
\end{align}
where $\bar{\rho}\equiv\sum_x l \sum_{\nu} \bar{\rho}_{\nu}(x)\approx M/L$ is the total mean density of particles, $\xi$ the coherence length, $\lambda_T$ the thermal wavelength and $k_{\rm max}$ the maximum momentum of excitations that we want to resolve in the theory. Seen another way, the discretization of space introduces a momentum cutoff $\sim \pi/l$, which must be greater than all relevant momentum scales in the system and thus
\begin{align}
\pi/\xi,\pi/\lambda_T,k_{\rm max}\ll \pi/l\ll \pi\bar{\rho}.\label{quasicondmom}
\end{align}
Note that the first and second inequalities in Eq.\,(\ref{quasicondmom}) are equivalent to the weakly interacting ($\mu\ll\hbar^2\bar{\rho}^2/m_b$) and low temperature conditions ($k_B T\ll\hbar^2\bar{\rho}^2/m_b$), respectively. These are characteristic of the 1D quasi-BEC regime and allows for the coarse-graining procedure. Putting this all together, the reservoir Hamiltonian in Eq.\,(\ref{Hres}) can be consistently rewritten as
\begin{align}
H_{\rm res}=\sum_{\nu,\lambda}\sum_x l&\left[\sqrt{\rho_{\nu}}e^{-i\theta_{\nu}}H^{\rm SOC}_{\nu\lambda}e^{i\theta_{\lambda}}\sqrt{\rho_{\lambda}}\right.\nonumber\\
&\left.+\frac{g_{\nu\lambda}}{2}\left(\rho_{\nu}\rho_{\lambda}-\frac{\delta_{\nu\lambda}}{l}\rho_{\nu}\right)\right]
\label{Hreslat},
\end{align}
where the discrete spatial derivatives contained in $H^{\rm SOC}_{\nu\lambda}$ are defined as $\Delta^2f/\Delta x^2\equiv[f(x+l)+f(x-l)-2f(x)]/l^2$ and $\Delta f/\Delta x\equiv [f(x+l)-f(x-l)]/[2l]$, with $f(x)$ an arbitrary function \cite{Mora:2003kbSM}. Importantly, the commutation relations for density and phase are also discretized and read
\begin{align}
[\rho_{\nu}(x),\theta_{\lambda}(x')]=i\delta_{\nu\lambda}\delta_{xx'}/l.\label{discretecomm}
\end{align}

\subsection{Perturbative expansion and diagonalization of the reservoir Hamiltonian}

We are now in position to identify the small parameters of the theory in order to perform a perturbative expansion of the Hamiltonian in Eq.\,(\ref{Hreslat}). A direct application of the method in Ref.\,\cite{Mora:2003kbSM} to our two-component case requires that, for each spin component separately, the relative density fluctuations and the phase fluctuation change over cells are small:
\begin{align}
\epsilon^{\nu}_1&\equiv |\delta\rho_{\nu}|/\bar{\rho}_{\nu}\ll 1,\label{e1}\\
\epsilon^{\nu}_2&\equiv l|\Delta\delta\theta_{\nu}/\Delta x|\ll 1.\label{e2}
\end{align}
Here $|A|$ represents the typical value of an operator $A$ in the state of the system. As in Ref.\,\cite{Whitlock:2003ixSM}, we additionally require that the difference between phase fluctuations in spin up and down components is small 
\begin{align}
\epsilon_3&\equiv |\delta\theta_{\uparrow}-\delta\theta_{\downarrow}|\ll 1.\label{e3}
\end{align}
In Sec.\,\ref{validity} we self-consistently check \emph{a posteriori} under which parameter conditions the assumptions (\ref{e1})-(\ref{e3}) indeed hold true, but for now we expand the reservoir Hamiltonian in Eq.\,(\ref{Hreslat}) up to second order in powers of the five small parameters $\epsilon$ as $H_{\rm res}\!=\!H^{(0)}_{\rm res}+H^{(1)}_{\rm res}+H^{(2)}_{\rm res}+{\cal O}(\epsilon^3)$. 

The zeroth order contribution can be written as
\begin{align}
H^{(0)}_{\rm res}=\sum_x l&\left[\frac{-\hbar^2}{2m_b}\sum_{\nu}\sqrt{\bar{\rho}_{\nu}}\frac{\Delta^2\sqrt{\bar{\rho}_{\nu}}}{\Delta x^2}+\sum_{\nu}(\nu\hbar\delta_0-\mu)\bar{\rho}_{\nu}\right.\nonumber\\
&+2\hbar\Omega_0 \cos(\bar{\Theta})\sqrt{\bar{\rho}_{\uparrow}\bar{\rho}_{\downarrow}}
+\frac{1}{2}\sum_{\nu,\lambda}g_{\nu\lambda}\bar{\rho}_{\nu}\bar{\rho}_{\lambda}\nonumber\\
&\left.+\frac{\hbar^2}{2m_b}\sum_{\nu}\bar{\rho}_{\nu}\left(\frac{\Delta \bar{\theta}_{\nu}}{\Delta x}-\nu k_0\right)^2\right],\label{H0}
\end{align}
where $\bar{\Theta}\!\equiv \!\bar{\theta}_{\uparrow}-\bar{\theta}_{\downarrow}$ is the zeroth order phase difference between spin components and the numerical values $\nu\!=\!\{+1,-1\}$ are assigned corresponding to $\nu\!=\!\{\uparrow,\downarrow\}$, respectively. The mean density and phase functions are determined by minimizing the energy functional $H^{(0)}_{\rm res}=H^{(0)}_{\rm res}[\bar{\rho}_{\nu},\bar{\theta}_{\nu}]$, yielding Gross-Pitaevskii (GP) type equations:
\begin{align}
&\frac{-\hbar^2}{2m_b}\frac{\Delta^2\sqrt{\bar{\rho}_{\nu}}}{\Delta x^2}+\frac{\hbar^2\sqrt{\bar{\rho}_{\nu}}}{2m_b}\left(\!\frac{\Delta\bar{\theta}_{\nu}}{\Delta x}\!-\!\nu k_0\!\right)^2\!\!+(\nu\hbar\delta_0-\mu)\sqrt{\bar{\rho}_{\nu}}\nonumber\\
&+\hbar\Omega_0\cos(\bar{\Theta})\sqrt{\bar{\rho}_{-\nu}}
+\sum_{\lambda}g_{\nu\lambda}\bar{\rho}_{\lambda}\sqrt{\bar{\rho}_{\nu}}=0,\label{GPrho}\\
&\frac{\hbar^2}{2m_b}\!\frac{\Delta}{\Delta x}\!\left[\bar{\rho}_{\nu}\!\left(\!\frac{\Delta\bar{\theta}_{\nu}}{\Delta x}\!-\!\nu k_0\!\right)\!\right]
\!\!+\!\hbar\Omega_0\nu\sin(\bar{\Theta})\sqrt{\bar{\rho}_{\uparrow}\bar{\rho}_{\downarrow}}\!=\!0.\label{GPtheta}
\end{align} 
Assuming that $\bar{\rho}_{\nu}(x)$ and $\bar{\theta}_{\nu}(x)$ are solutions of these GP equations, one can show that the first order correction of $H_{\rm res}$ vanishes exactly $H^{(1)}_{\rm res}=0$, as it is also the case for the single-component quasi-BEC treatment \cite{Mora:2003kbSM}. To make use of known analytical solutions for $\bar{\rho}_{\nu}$ and $\bar{\theta}_{\nu}$, it is convenient to define the complex classical field $\bar{\psi}_{\nu}\equiv e^{i\bar{\theta}_{\nu}}\sqrt{\bar{\rho}_{\nu}}$, such that $H^{(0)}_{\rm res}$ in Eq.\,(\ref{H0}) can be rewritten as
\begin{align}
H^{(0)}_{\rm res}=\sum_x l\sum_{\nu,\lambda}\left[\bar{\psi}^{\ast}_{\nu}H^{\rm SOC}_{\nu\lambda}\bar{\psi}_{\lambda}+\frac{g_{\nu\lambda}}{2}|\bar{\psi}_{\nu}|^2|\bar{\psi}_{\lambda}|^2\right].\label{Hreszeropsi}
\end{align}
Taking the continuum limit, Eq.\,(\ref{Hreszeropsi}) is formally the same mean-field energy functional used in Ref.\,\cite{Li:2012ddSM} to predict a rich phase diagram for the homogeneous 3D BEC with SOC. Different phases were found as a function of $\Omega_0$ and the three interaction parameters $G_1\!\equiv\!(\bar{\rho}/8)(g_{\uparrow\uparrow}+g_{\downarrow\downarrow}+2g_{\uparrow\downarrow})$, $G_2\!\equiv\!(\bar{\rho}/8)(g_{\uparrow\uparrow}+g_{\downarrow\downarrow}-2g_{\uparrow\downarrow})$ and $G_3\!\equiv\!(\bar{\rho}/4)(g_{\uparrow\uparrow}-g_{\downarrow\downarrow})$, when keeping the detuning fixed to $\hbar\delta_0=-G_3$. Nevertheless, for our reservoir engineering purposes with the 1D SOC quasi-BEC, we are interested in the particular situation where quasi-condensation occurs deterministically at a finite positive wavenumber $k_m\!>\!0$, for all values of $\Omega_0$. Using the same variational approach as in Ref.\,\cite{Li:2012ddSM}, it can be shown that when having a finite negative detuning $\delta_0<0$ satisfying
\begin{align}
2G_2+G_3<\hbar|\delta_0|\ll E_0,\label{detuningcond}
\end{align} 
the required magnetized phase can always be prepared (assuming $G_1\!>\!0$). In the experimentally relevant case of $^{87}$Rb, the conditions (\ref{detuningcond}) is particularly easy to meet, because the interaction parameters satisfy $0<2G_2=G_3\ll G_1,E_0$. Putting all of this together, the zeroth order solutions for density, phase and ground state energy read
\begin{align}
\bar{\rho}_{\nu}&=\bar{\rho}(1+\nu q)/2,\label{zerothdensity}\\
\bar{\theta}_{\nu}&=qk_0x+\pi(\nu-1)/2,\label{zerothphase}\\
E_{\rm GS}&=M\left[E_0+G_1-q(\hbar|\delta_0|-G_3)\right.\nonumber\\
&\hspace{1cm}\left.-q^2(E_0-G_2)-\hbar\Omega_0\sqrt{1-q^2}\right].\label{zerothenergy}
\end{align}
Here $q\equiv k_m/k_0\in[0,1]$ corresponds to the only positive solution of the 4th order equation
\begin{align}
q^4+2Cq^3+(C^2+D^2-1)q^2-2Cq-C^2=0,
\end{align}
with 
\begin{align}
C&\equiv\frac{\hbar|\delta_0|-G_3}{2(E_0-G_2)}\ll 1,\\
D&\equiv \frac{\hbar\Omega_0}{2(E_0-G_2)}.
\end{align} 
For $D\ll1$, we obtain up to second order in $C$ and $D$, $q=1-D^2/2\lesssim 1$ and for $D\gtrsim 1$, $q$ approaches zero, but never vanishes exactly if $C\neq 0$. The chemical potential $\mu\equiv\partial E_{\rm GS}/\partial N$ is obtained directly from Eq.\,(\ref{zerothenergy}), which in the limit $D\ll1$ takes the simple form $\mu=\bar{\rho}g_{\uparrow\uparrow}-\hbar|\delta_0|-E_0D^2$, again up to second order in $C$ and $D$. 

\begin{figure}[ht!]
\includegraphics[angle=0,width=0.48\textwidth]{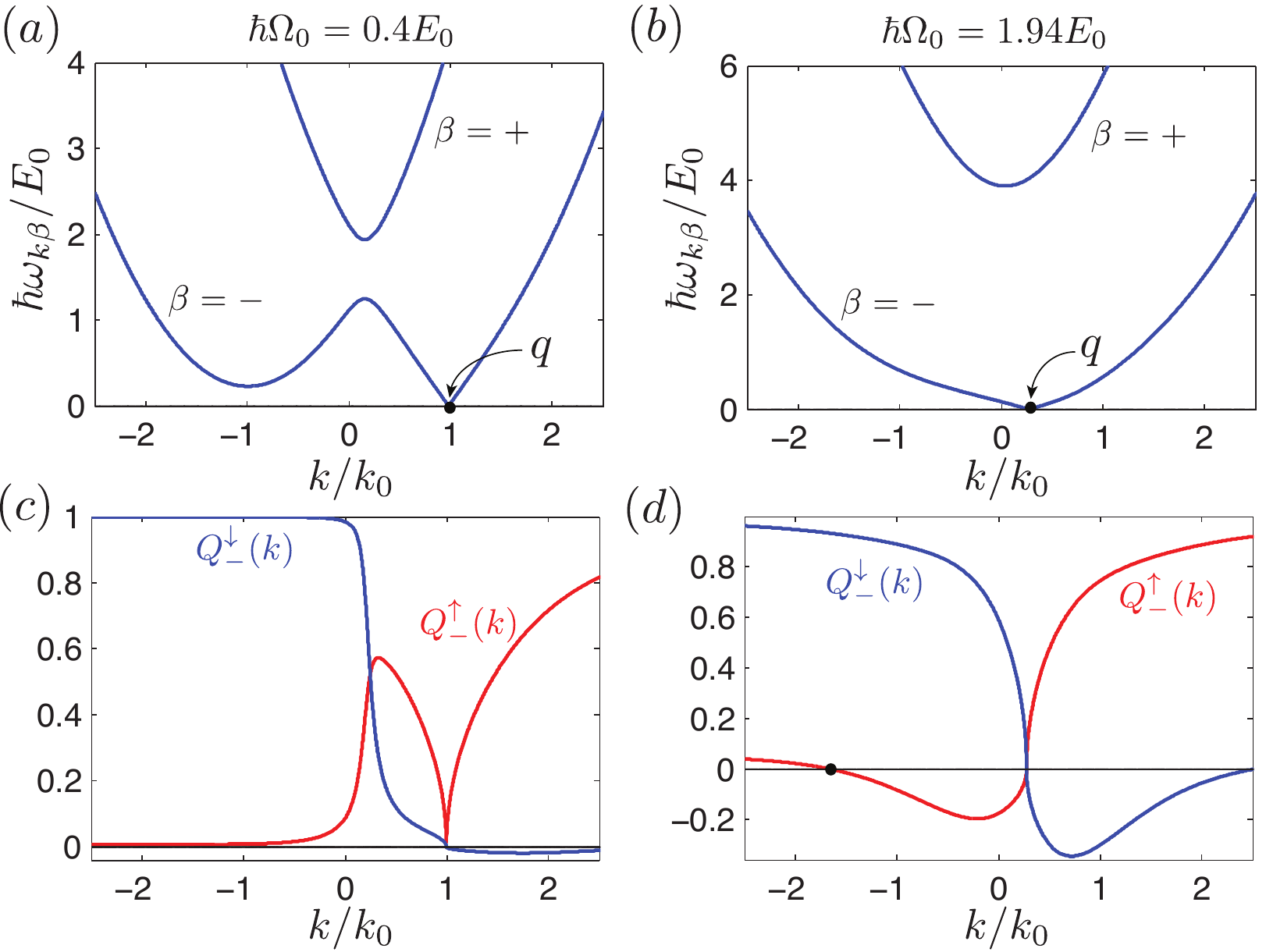}
\caption{Numerical excitation spectra $\hbar\omega_{k\beta=\pm}$ and density fluctuation coefficients $Q^{\nu}_{-}(k)$ (with $\nu=\uparrow,\downarrow$) in the plane-wave phase (with quasi-condensation at $k_m=qk_0$), for $\hbar\Omega_0\!=\!0.4E_0$, $\hbar\delta_0\!=\!-0.1E_0$ (a,c) and $\hbar\Omega_0\!=\!1.94E_0$, $\hbar\delta_0\!=\!-0.004E_0$ (b,d). Other parameters are $\bar{\rho}\!=\!6.14k_0$ and $g_{\uparrow\uparrow}\!=\!g_{\uparrow\downarrow}\!=\!g_{\downarrow\downarrow}\!=\!0.23E_0/k_0$.}
\label{fig:spectrum_Qs}
\end{figure}

As in Ref.\,\cite{Mora:2003kbSM}, we use the second order correction of the Hamiltonian $H_{\rm res}^{(2)}$ to calculate the Heisenberg equations of motion for the density and phase fluctuations. In our particular case, assuming the zeroth order solution in Eqs.\,(\ref{zerothdensity})-(\ref{zerothphase}), they read
\begin{align}
\hbar\delta\dot{\rho}_{\nu}&=-\frac{\hbar^2\bar{\rho}_{\nu}}{m_b}\frac{\Delta^2\delta\theta_{\nu}}{\Delta x^2}-\frac{\hbar^2k_0}{m_b}(q-\nu)\frac{\Delta\delta\rho_{\nu}}{\Delta x}\nonumber\\
&+2\hbar\Omega_0\sqrt{\bar{\rho}_{\uparrow}\bar{\rho}_{\downarrow}}(\delta\theta_{\nu}-\delta\theta_{-\nu}),\label{Heisenbergden}\\
\hbar\delta\dot{\theta}_{\nu}&=\frac{\hbar^2}{4m_b\bar{\rho}_{\nu}}\frac{\Delta^2\delta\rho_{\nu}}{\Delta x^2}-\frac{\hbar^2k_0}{m_b}(q-\nu)\frac{\Delta\delta\theta_{\nu}}{\Delta x}-\sum_{\lambda}g_{\nu\lambda}\delta\rho_{\lambda}\nonumber\\
&-\frac{\hbar\Omega_0}{2\bar{\rho}_{\nu}}\left(\sqrt{\frac{\bar{\rho}_{-\nu}}{\bar{\rho}_{\nu}}}\delta\rho_{\nu}-\sqrt{\frac{\bar{\rho}_{\nu}}{\bar{\rho}_{-\nu}}}\delta\rho_{-\nu}\right).\label{Heisenbergphase}
\end{align}
It is convenient to define the non-hermitian operators
\begin{align}
B_{\nu}(x)\equiv \frac{\delta\rho_{\nu}}{2\sqrt{\bar{\rho}_{\nu}}}+i\sqrt{\bar{\rho}_{\nu}}\delta\theta_{\nu},\label{nonhermitian}
\end{align}
which by construction obey bosonic commutation relations $[B_{\nu}(x),B_{\lambda}^{\dag}(x')]=\delta_{\nu\lambda}\delta_{x,x'}/l$ and whose dynamics is governed by the linear equations
\begin{align}
i\hbar\dot{B}_{\nu}\!={\cal L}_{\nu}B_{\nu}\!+\!(\eta_{\uparrow\downarrow}\!-\!\hbar\Omega_0)B_{-\nu}\!+\!\eta_{\nu\nu}B_{\nu}^{\dag}+\eta_{\uparrow\downarrow} B_{-\nu}^{\dag},\label{BdG}
\end{align}
with
\begin{align}
{\cal L}_{\nu}&\!\equiv\!\frac{-\hbar^2}{2m_b}\!\frac{\Delta^2}{\Delta x^2}\!-\!\frac{i\hbar^2 k_0}{m_b}\!(q\!-\!\nu)\!\frac{\Delta}{\Delta x}\!+\!\hbar\Omega_0\!\sqrt{\frac{\bar{\rho}_{-\nu}}{\bar{\rho}_{\nu}}}\!+\!\eta_{\nu\nu},\\
\eta_{\nu\lambda}&\equiv g_{\nu\lambda}\sqrt{\bar{\rho}_{\nu}\bar{\rho}_{\lambda}}.
\end{align}
Importantly, in terms of these bosonic operators $B_{\nu}$, the second order Hamiltonian $H_{\rm res}^{(2)}$ takes the quadratic form
\begin{align}
H^{(2)}_{\rm res}\!=\!&\sum_x\! l\sum_{\nu}\!\left[\!B_{\nu}^{\dag}{\cal L}_{\nu}B_{\nu}\!+\!\frac{\!(\eta_{\uparrow\downarrow}\!-\!\hbar\Omega_0)}{2}(B_{\nu}^{\dag}B_{-\nu}\!+\!B_{-\nu}^{\dag}B_{\nu})\right.\nonumber\\
&\left.+\frac{\eta_{\nu\nu}}{2}(B_{\nu}^{\dag}{}^2\!+\!B_{\nu}^2)+\frac{\eta_{\uparrow\downarrow}}{2}(B_{\nu}^{\dag}B_{-\nu}^{\dag}\!+\!B_{-\nu}B_{\nu})\right],
\end{align}
which can be straightforwardly diagonalized using the standard Bogoliubov-de-Gennes (BdG) procedure \cite{Mora:2003kbSM,PitaevskiiStringariSM}. In the present homogeneous case, the normal mode decomposition of $B_\nu$ in the Heisenberg picture reads
\begin{align}
B_{\nu}(t)&=\frac{1}{\sqrt{L}}\sum_{k,\beta}u_{k\beta}^{\nu}b_{k\beta}e^{i[(k-k_m)x-\omega_{k\beta}t]}\nonumber\\
&+\frac{1}{\sqrt{L}}\sum_{k,\beta}v_{k\beta}^{\nu}b_{k\beta}^{\dag}e^{-i[(k-k_m)x-\omega_{k\beta}t]},\label{ansatz}
\end{align}
where the real coefficients $u_{k\beta}^{\nu}$ and $v_{k\beta}^{\nu}$ [normalized as $\sum_{\nu}(u_{k\beta}^{\nu})^2\!\!-\!(v_{k\beta}^{\nu})^2\!=\!\!1$], as well as the the excitation spectrum $\omega_{k\beta}$, are determined from the BdG equations arising from Eq.\,(\ref{BdG}). As a result, the reservoir Hamiltonian up to second order in the five small parameters $\epsilon$, takes the diagonal form
\begin{align}
H_{{\rm res}}=E_{{\rm GS}}+\sum_{k,\beta}\hbar\omega_{k\beta}b_{k\beta}^{\dag}b_{k\beta}+{\cal O}(\epsilon^3).\label{HresDiag}
\end{align}
Here the bosonic Bogoliubov-like operators $b_{k\beta}$, satisfying $[b_{k\beta},b_{k'\beta'}^{\dag}]=\delta_{kk'}\delta_{\beta\beta'}$, annihilate an elementary excitation with wavenumber $k$ and at branch $\beta=\pm$. In Fig.\,\ref{fig:spectrum_Qs}(a-b) we show typical excitation spectra $\hbar\omega_{k\beta=\pm}$ for the 1D quasi-BEC prepared in the plane-wave phase. We note that Eq.\,(\ref{HresDiag}) is the expression for the reservoir Hamiltonian used in the main text, where the ground state energy $E_{{\rm GS}}$ has been omitted. Using Eqs.\,(\ref{nonhermitian}) and (\ref{ansatz}), the density fluctuation operator in the Schr\"odinger picture can be conveniently expressed in terms of the elementary excitations as, 
\begin{align}
\delta\rho_{\nu}\!&\!=\sqrt{\frac{\bar{\rho}_{\nu}}{L}}\sum_{k,\beta}\!Q_{\beta}^{\nu}(k)[b_{k\beta}e^{i(k-k_m)x}+{\rm h.c.}],\label{densityfluctuation}
\end{align}
where the coefficients $Q_{\beta}^{\nu}(k)\equiv u_{k\beta}^{\nu}+v_{k\beta}^{\nu}$ reflect the strong spin polarization of the excitations with $|k|\gtrsim k_0$ [cf.~Fig.\,\ref{fig:spectrum_Qs}(c-d)]. Note that for the parameters in Fig.\,\ref{fig:spectrum_Qs}(d), $Q^{\uparrow}_{-}(k)$ has a zero crossing at a negative wave vector $k$. This allows for the realization of an ideal unidirectional reservoir as mentioned in the main text.

\subsection{Validity of the expansion}\label{validity}

Following Ref.\,\cite{Mora:2003kbSM}, the order of magnitude of the assumed small parameters in Eqs.\,(\ref{e1})-(\ref{e3}) can be self-consistently estimated by their root mean square values in a thermal state as $\epsilon^{\nu}_1\sim\langle\delta\rho_{\nu}^2/\bar{\rho}_{\nu}^2\rangle^{1/2}$, $\epsilon^{\nu}_2\sim l\langle\left(\Delta\delta\theta_{\nu}/\Delta x\right)^2\rangle^{1/2}$ and $\epsilon_3\sim\langle(\delta\theta_{\uparrow}-\delta\theta_{\downarrow})^2\rangle^{1/2}$. Using Eq.\,(\ref{nonhermitian}), the solution of the BdG equations (\ref{ansatz}) and taking the continuum limit in the sum over $k$, the corresponding expectation values read
\begin{align}
\left\langle\frac{\delta\rho_{\nu}^2}{\bar{\rho}_{\nu}^2}\right\rangle\!&=\!\int\!\!\frac{dk}{2\pi\bar{\rho}_{\nu}}\!\sum_{\beta}(u_{k\beta}^{\nu}\!+\!v_{k\beta}^{\nu})^2(2n_{k\beta}\!+\!1),\label{e1int}\\
\left\langle\!\!\left(\!\frac{\Delta\delta\theta_{\nu}}{\Delta x}\!\right)^2\right\rangle\!&=\!\!\int\!\frac{k^2dk}{8\pi\bar{\rho}_{\nu}}\sum_{\beta}(u_{k\beta}^{\nu}\!-\!v_{k\beta}^{\nu})^2(2n_{k\beta}\!+\!1),\label{e2int}\\
\left\langle(\delta\theta_{\uparrow}\!\!-\!\delta\theta_{\downarrow})^2\right\rangle\!&=\!\!\int\!\frac{dk}{8\pi}\!\sum_\beta\!\!\left[\!\sum_{\nu}\!\frac{(u_{k\beta}^{\nu}\!\!-\!v_{k\beta}^{\nu})}{\nu\sqrt{\bar{\rho}_{\nu}}}\!\right]^2\!\!\!\!\!(2n_{k\beta}\!+\!1).\label{e3int}
\end{align}
Here the integrals run over the domain $k\in[-\pi/l,\pi/l]$ and $n_{k\beta}\equiv 1/(e^{\hbar\omega_{k\beta}/k_BT}-1)$ is the usual Bose distribution. Integrating numerically Eqs.\,(\ref{e1int})-(\ref{e3int}), one can show that provided the inequalities in Eq.\,(\ref{quasicondlength}) hold (with $k_{\rm max}\sim k_0$), the $\epsilon$ parameters are of order
\begin{align}
\epsilon^{\nu}_1&\sim\epsilon^{\nu}_2\sim\frac{1}{\sqrt{\bar{\rho}_{\nu}l}}\sim\frac{1}{\sqrt{\bar{\rho}l}}\frac{1}{\sqrt{1+\nu q}},\\
\epsilon_3&\sim\frac{1}{\sqrt{\bar{\rho}_{\downarrow}l}}\sim\frac{1}{\sqrt{\bar{\rho}l}}\frac{1}{\sqrt{1-q}}.
\end{align}
For $\bar{\rho}l\!\gg\!1$, as assumed in Eq.\,(\ref{quasicondlength}), all these parameters are small except in the limit $q\approx 1$ ($\hbar\Omega_0\ll E_0$), where there is strong spin polarization in the quasi-BEC along the $\ket{\uparrow}$ state ($\bar{\rho}_{\uparrow}\gg\bar{\rho}_{\downarrow}$). To also capture this parameter regime in our theory, we present in the following a slightly more general approach.

\subsection{Expansion in a rotated spin basis}\label{rotatedbasis}

We change the reference frame by applying a global spin rotation around the $y$ axis, $R\equiv e^{i(\theta_q/2)\tau_y}$. The transformed field operators read
\begin{align}
\psi_{+}(x)&\equiv\cos(\theta_q/2)\psi_{\uparrow}+\sin(\theta_q/2)\psi_{\downarrow},\\
\psi_{-}(x)&\equiv-\sin(\theta_q/2)\psi_{\uparrow}+\cos(\theta_q/2)\psi_{\downarrow},
\end{align}
which can be further expressed in the density-phase representation as $\psi_{\alpha}(x)\!\equiv\!e^{i(\bar{\theta}_{\alpha}+\delta\theta_{\alpha})}\sqrt{\bar{\rho}_{\alpha}
+\delta\rho_{\alpha}}$, with $\alpha\!=\!\{+,-\}$. Here, $\delta\rho_{\alpha}$ and $\delta\theta_{\alpha}$ denote the density and phase fluctuations of the reservoir atoms in spin states $\ket{\alpha}\!=\!\{\ket{+},\ket{-}\}$, around the mean values $\bar{\rho}_{\alpha}$ and $\bar{\theta}_{\alpha}$, respectively. The rotation angle $\theta_q\equiv \arctan(q/\sqrt{1-q^2})$ is chosen such that, for all values of $q$, the quasi-BEC atoms have equal populations in both spin states $\ket{\alpha}$ and therefore equal zeroth order densities $\bar{\rho}_{\alpha}=\bar{\rho}/2$. In this rotated basis, we perform exactly the same Mora-Castin discretization and expansion procedure as done above in the $\lbrace\uparrow,\downarrow\}$ basis. As a result, the BdG equations for the bosonic operators $B_{\alpha}\equiv \delta\rho_{\alpha}/(2\sqrt{\bar{\rho}_{\alpha}})+i\delta\theta_{\alpha}$ read
\begin{align}
i\hbar\dot{B}_{\alpha}\!={\cal L}_{\alpha}B_{\alpha}+{\cal P}B_{-\alpha}+(g\bar{\rho}/2)(B_{\alpha}^{\dag}+B_{-\alpha}^{\dag}),\label{BdG2}
\end{align}
where
\begin{align}
{\cal L}_{\alpha}&\!\equiv\!\frac{-\hbar^2}{2m_b}\!\frac{\Delta^2}{\Delta x^2}\!-\!\frac{i\hbar^2 k_0}{m_b}\left[q-\alpha\cos(\theta_q)\right]\!\frac{\Delta}{\Delta x}\!+\!\frac{g\bar{\rho}}{2}\nonumber\\
&+\cos(\theta_q)\hbar\Omega_0+\sin(\theta_q)\left(\frac{\hbar^2k_0^2q}{m_b}+\hbar|\delta_0|\right),\\
{\cal P}&\equiv\!\frac{i\hbar^2 k_0}{m_b}\sin(\theta_q)\frac{\Delta}{\Delta x}\!-\cos(\theta_q)\hbar\Omega_0\nonumber\\
&-\!\sin(\theta_q)\left(\frac{\hbar^2k_0^2q}{m_b}\!+\!\hbar|\delta_0|\right)\!+\!\frac{g\bar{\rho}}{2}.
\end{align}
For notational simplicity we assume the intraspecies coupling constants to be all equal $g_{\uparrow\uparrow}=g_{\downarrow\downarrow}=g_{\uparrow\downarrow}\equiv g$. The solution to the BdG equations (\ref{BdG2}) can be again expressed in terms of the Bogoliubov-like excitations as
\begin{align}
B_{\alpha}(t)&=\frac{1}{\sqrt{L}}\sum_{k,\beta}u_{k\beta}^{\alpha}b_{k\beta}e^{i[(k-k_m)x-\omega_{k\beta}t]}\nonumber\\
&+\frac{1}{\sqrt{L}}\sum_{k,\beta}v_{k\beta}^{\alpha}b_{k\beta}^{\dag}e^{-i[(k-k_m)x-\omega_{k\beta}t]},\label{ansatz2}
\end{align}
where the real coefficients $u_{k\beta}^{\alpha}$ and $v_{k\beta}^{\alpha}$ are normalized as $\sum_{\alpha}(u_{k\beta}^{\alpha})^2\!\!-\!(v_{k\beta}^{\alpha})^2\!=\!\!1$ and the dispersion relation $\omega_{k\beta}$ is the same as above. The advantage of this basis is that the expansion parameters, calculated analogously to Eqs.\,(\ref{e1int})-(\ref{e3int}) and under the same assumptions of Eq.\,(\ref{quasicondlength}), are always small independent on the value of $q$: $\epsilon_1^{\alpha}\approx\epsilon_2^{\alpha}\approx 1/\sqrt{\rho_{\alpha}l}\approx 1/\sqrt{\bar{\rho}l/2}\ll 1$ and $\epsilon_3\approx 1/\sqrt{\bar{\rho}l}\ll 1$. Finally, the density fluctuations $\delta\rho_{\nu}$ in the original $\{\ket{\uparrow},\ket{\downarrow}\}$ spin states can always be expressed in terms of the elementary excitations using Eq.\,(\ref{densityfluctuation}), with the spinor coefficients given by
\begin{align}
Q_{\beta}^{\nu}(k)\!=\!\sum_{\alpha}\!\left[\frac{1+\nu\sin(\theta_q)+\nu\alpha\cos(\theta_q)}{2\sqrt{1+\nu q}}\right]\!\!(u_{k\beta}^{\alpha}+v_{k\beta}^{\alpha}).
\end{align}
We note that even though the approximations are justified only in the rotated basis for all values of $q$, the final results are independent of the basis choice. 

\section{System-reservoir interaction}

In this section we make use of the density fluctuation expression in Eq.\,(\ref{densityfluctuation}), to derive the system-reservoir interaction Hamiltonian $H_{\rm int}$, given in Eq.\,(3) of the main text. We also comment on the lattice wavevector constraint required in our implementation, as well as on the inclusion of a finite trapping potential for atoms $b$ (aligned with the optical lattice).

\subsection{Derivation of interaction Hamiltonian and decay rates}

On a microscopic level, the undriven system Hamiltonian for the lattice atoms $a$ (with mass $m_a$) is given by
\begin{align}
H_{\rm sys}^{(0)}=\!\int\! dx\, \psi_a^{\dag}\!\left(-\frac{\hbar^2}{2m_a}\frac{d^2}{dx^2}+V(x)\right)\!\psi_a+\frac{g_a}{2}{\psi_a^\dag}\psi_a^\dag\psi_a\psi_a,
\end{align}
where $\psi_a(x)$ is the field operator of atomic species $a$, trapped in a 1D optical lattice potential $V(x)=V_0\sin^2(\pi x/d)$ of period $d$, and $g_a$ is the 1D interaction constant. To map this system to a 1D chain of spins we consider a situation, where the lattice filling is one atom per lattice site, and the lattice is deep [$V_0\gg \hbar^2k_{\rm lat}^2/(2m_a)$, where $k_{\rm lat}\equiv \pi/d$\,], such that tunnelling between the sites is suppressed. Restricting the dynamics to the two lowest vibrational states on each site $j$, i.e. the Wannier states $w_g(x-x_j)$ and $w_e(x-x_j)$, the mapping to two-level systems (TLSs) is formally achieved by replacing $\psi_a(x)\rightarrow\sum_j w_g(x-x_j) \ket{g}_j+\sum_j w_e(x-x_j) \ket{e}_j$, giving the undriven system Hamiltonian $H^{(0)}_{\rm sys}=\hbar\omega\sum_j \ket{e}_j\bra{e}$ (up to an irrelevant constant). In addition these TLSs can be driven by coupling the two lowest vibrational states via a Raman process, leading to the final system Hamiltonian in Eq.\,(2) of the main text.
As with the intraspecies interactions, the interspecies interactions on a microscopical level stem from $s$-wave collisions between the system atoms $a$ and the reservoir atoms $b$. The Hamiltonian accounting for this system-reservoir interaction can be written as 
\begin{align}
H_{{\rm int}}=\sum_{\nu=\uparrow,\downarrow}g_{a\nu}\int \!dx\ \psi_a^\dag\psi_a\psi_\nu^\dag\psi_\nu,\label{Hinteract}
\end{align}
where $g_{a\nu}$ are the effective 1D interspecies interaction constants. We note that, in writing Eq.\,\eqref{Hinteract} we exclude the possibility for spin changing collisions. This can be achieved by a suitable choice of the atomic species $a$ with zero electronic angular momentum, e.g.~Ytterbium, such that spin changing collisions are prohibited by angular momentum conservation. In this interaction Hamiltonian (\ref{Hinteract}) we can replace the density of system atoms $a$ as
\begin{align}
&\psi_a^{\dag}(x)\psi_a(x)\rightarrow\sum_j\sum_{r=e,g} |w_r(x-x_j)|^2 \ket{r}_j\bra{r}\nonumber\\
&+\sum_j \left(w_{e}(x-x_j)w_{g}(x-x_j)\ket{e}_j\bra{g}+\rm h.c.\right),
\end{align}  
and use $\psi_\nu^\dag\psi_\nu\equiv\bar\rho_\nu+\delta\rho_\nu$, with the density fluctuations of reservoir atoms given in Eq.\,(\ref{densityfluctuation}). We note that the system atoms couple only to the density fluctuations in the reservoir. As a result, one can express $H_{\rm int}$ in terms of elementary excitations as
\begin{align}
H_{{\rm int}}=&\sum_{k,\beta,j}\sum_{r,r'=g,e}\!\!G_{\beta}^{r,r'}(k)\ket{r}_j\bra{r'}b_{k\beta}e^{i(k-k_m)x_j}+\rm h.c.,\label{intfirst}
\end{align}
with the coupling constants
\begin{align}
G_{\beta}^{r,r'}(k)&\!=\!\sum_{\nu}g_{a\nu}\sqrt{\frac{\bar{\rho}_{\nu}}{L}}Q_{\beta}^{\nu}(k)\!\int\! dx\,w_{r}(x)w_{r'}(x)e^{i(k-k_m)x}.\label{coupling}
\end{align}
In a rotating-wave approximation (RWA) \cite{QuantumNoiseSM} we neglect intraband couplings, because (\emph{i}) the coupling constants $G^{r,r'}_{-}(k)\rightarrow 0$ for $k\rightarrow k_m$ reflecting the vanishing static structure factor in the phononic part of the Bogoliubov spectrum, and (\emph{ii}) the roton gap (at $k_{\rm rot}\!\sim\!-k_m$) is large enough such that excitations around the roton minimum are suppressed, i.e. $\hbar\omega_{k_{\rm rot},-}\gg |G_{-}^{r,r}(k_{\rm rot})|$ [cf.~Fig.\,\ref{fig:spectrum_Qs}(a)]. In this RWA, we can therefore restrict the reservoir only to resonant excitations with energies around the interband transition frequency $\omega$. By placing $\hbar\omega$ in the spin-orbit gap at energies $\sim E_0$ [cf.~Fig.1(d) of the main text] there are two such resonant types of excitations, left moving ones with wavevectors $k\in [k_{L}-k_{\theta},k_{L}+k_{\theta}]$ and group velocity $v_L<0$, and right moving ones with wavevectors $k\in [k_{R}-k_{\theta},k_{R}+k_{\theta}]$ and group velocity $v_R>0$. Here $k_{\theta}$ is a momentum cutoff due to the RWA \cite{QuantumNoiseSM}. This allows us to write the system reservoir interaction Hamiltonian in the form
\begin{align}
H_{{\rm int}}=&i\hbar\!\!\sum_{s=L,R}\!\sqrt{\frac{\gamma_{s}|v_{s}|}{L}}\!\sum_{k=k_{s}-k_{\theta}}^{k_{s}+k_{\theta}}\!\!\sum_{j}\!\sigma_{j}^{\dag}b_{k,-}e^{i(k-k_m)x_{j}}+{\rm h.c.},
\end{align}
with couplings to left and right moving modes,
\begin{align}
\gamma_{s}\equiv\frac{\eta(k_s)e^{-\eta(k_s)}}{\hbar^{2}|v_{s}|}\!\left(\sum_{\nu}g_{a\nu}\sqrt{\bar{\rho}_\nu}Q_{-}^{\nu}(k_s)\right)^2,\label{decayratess}
\end{align} 
also given in Eqs.\,(3) and (4) of the main text ($s=L,R$). Here we evaluated the integrals in Eq.\,\eqref{coupling}, by approximating the Wannier states with harmonic oscillator wavefunctions giving $\eta(k)\equiv(E_0/\hbar\omega)(m_b/m_a)[(k-k_m)/k_0]^2$.

\subsection{Constraint on the lattice wavevector}

We note that the requirement $\hbar\omega\!\sim\!E_0$ constrains the choice of lattice depth $V_0$ and lattice wave vector $k_{\rm lat}$. Taking this into account, the condition for a deep lattice required to define our two-level system reads, 
\begin{align}
\left(\frac{2E_0}{\hbar\omega}\right)^2\left(\frac{m_b}{m_a}\right)^2
\left(\frac{k_{\rm lat}}{k_0}\right)^4\ll 1.
\end{align}
To satisfy this condition, it is required that the lattice wavevector is smaller than the Raman one, $k_0>k_{\rm lat}$ and also that the lattice atoms are heavier than the reservoir ones, $m_a>m_b$.

\subsection{Inclusion of a shallow trapping potential aligned with the optical lattice}

Throughout this work we have assumed the reservoir to be perfectly infinite and homogeneous along one dimension. In this case, the system-bath coupling constants in Eq.\,(\ref{coupling}) are independent of $j$ and the resulting decay rates to left and right moving modes in Eq.(\ref{decayratess}) are homogeneous. The inclusion of a finite trapping potential will introduce inhomogeneities in the system (e.~g. in the density $\bar{\rho}$), as well as boundaries for the propagation of the reservoir excitations (similar to mirrors). The validity of the Markovian master equation in Eq.\,(5) of the main text relies on the fact that the reservoir excitations propagate out to infinity, and therefore the trapping depth must be lower than the energy $\hbar\omega$ of these excitations, such that they can actually escape from the trap \cite{{Chen:2014itSM}}. Secondly, it is possible to generalize our master equation to account for inhomogeneous couplings, as shown in Eq.\,(37) of Ref.\,\cite{Stannigel:2011igSM}. The dimer formation is not altered by these inhomogeneities, as long as the density $\bar{\rho}(x)$ varies slowly on a scale $\sim 2d$, such that two neighboring spins can still couple equally to the chiral reservoir and locally dimerize.

\section{Dimerized steady state solution for asymmetric bidirectional coupling}

In this section we give a proof that the dimerised pure state $\ket{\psi}=\bigotimes_{i=1}^{N/2}\ket{D}_{2i-1,21}$ is a steady state of Eq.\,(5) of the main text for any ratio $0\leq\gamma_L/\gamma_R<1$. We also perform numerics for small system sizes in order to extract the scaling of the timescale $t_{\rm ss}$ to reach this state as $\gamma_L/\gamma_R\rightarrow 1$.

\subsection{Construction of the dimerized steady state}

We remind the reader that we consider the setting $\Omega_i=\Omega$, $\nu=\omega$ and $(k_R-k_L)d= 4\pi n$ (with $n$ integer). We additionally assume $\Omega^{\ast}=\Omega$ without loss of generality. In this case, the master equation can be rewritten as
\begin{align}
\dot\rho&=-(i/\hbar)[H_{\rm sys},\rho]+\mathcal{L}_L\rho+\mathcal{L}_R\rho,\label{mastershort}
\end{align}
where the Liouvillian term $\mathcal{L}_{R}$ describes cascaded evolution to the right, and $\mathcal{L}_L$ cascaded evolution to the left, that is
\begin{align}
\mathcal{L}_{L}\rho\!&\equiv\!\frac{\gamma_L}{2}\!\sum_{j}\!{\cal D}({\sigma}_{j},{\sigma}_{j})\rho+\!\gamma_L\sum_{j>l}\!\left([{\sigma}_{j},\rho{\sigma}_{l}^{\dag}]\!+\textrm{h.c.}\right)\!,\\
\mathcal{L}_{R}\rho\!&\equiv\!\frac{\gamma_R}{2}\!\sum_{j}\!{\cal D}({\sigma}_{j},{\sigma}_{j})\rho+\!\gamma_R\sum_{j<l}\!\left([{\sigma}_{j},\rho{\sigma}_{l}^{\dag}]\!+\textrm{h.c.}\right)\!.
\end{align}
For notational convenience we go to a rotating frame with the driving frequency, such that (with an abuse of notation) the system Hamiltonian becomes $H_{\rm sys}=\hbar\Omega \sum_j(\sigma_j+\sigma_j^{\dag})$.
To construct the steady state of this master equation, we first consider a purely unidirectional system, that is $\gamma_L=0$. In Ref.\,\cite{Stannigel:2012jkSM} it was shown that when driving the the latter system symmetrically and on resonance with Rabi frequency $\Omega_R$, the steady state is pure and a product of dimers $\ket{\Psi}=\bigotimes_{j=1}^{N/2}\ket{D}_{2j-1,2j}$, with $\ket{D}=(\ket{gg}+\alpha_R\ket{S})/\sqrt{1+|\alpha_R|^2}$, $\ket{S}\equiv\frac{1}{\sqrt{2}}(\ket{ge}-\ket{eg})$ with a singlet fraction $\alpha_R=2i\sqrt{2}\Omega_R/\gamma_R$. This can be shown by exploiting the unidirectional character of the master equation in the folowing sense: The unidirectionality leads to a closed equation for the reduced system density operator of the first two spins. This reduced equation has the dark state $\ket{D}$ as its unique, pure steady state. Since this state is pure, the first two spins are not entangled with the rest of the system and thus the first pair can be factorized out. Once this first dimer is formed, and the first two spins are in their dark state, the equation of motion for the third and fourth spin also decouple from the rest and these two spins are driven into the dimer state as well. This argument can be repeated iteratively to show that indeed the product of such dimers is the unique steady state of a purely unidirectional cascaded spin chain with an even number of spins. Completely analogously, one can show that the unique steady state of a system (with an even number of spins) that is cascaded in the opposite direction ($\gamma_R=0$), is also a product of dimers, but the sign of $\alpha$ is reversed ($\alpha_L=-i\sqrt{2}\Omega_L/\gamma_L$) since the singlet is anti-symmertric under exchange of the spins. 

If we now consider asymmetric bidirectional decay in the system, i.e. a cascaded channel to the left and to the right ($\gamma_L,\gamma_R>0$) with different strengths $\gamma_L\neq\gamma_R$, such an iterative solution for the steady state is no longer possible due to the lack of strict unidirectionality. However, the master equation (\ref{mastershort}) still has a dimerized unique dark steady state for any nonzero value of the decay asymmetry $\Delta\gamma$. To identify the steady state, we split the system Hamiltonian as 
\begin{align}
H_{\rm sys}=\gamma_R\frac{\hbar\Omega}{\Delta\gamma}\sum_j(\sigma_j+\sigma_j^{\dag})-\gamma_L\frac{\hbar\Omega}{\Delta\gamma} \sum_j(\sigma_j+\sigma_j^{\dag}),
\end{align}
so that we can write the total master equation as a sum of a cascaded one to the right that is driven with a Rabi frequency $\Omega_R\equiv\gamma_R\Omega/\Delta\gamma$ and a cascaded one to the left, driven with  Rabi frequency $\Omega_L\equiv-\gamma_L\Omega/\Delta\gamma$. From the above discussion we know that both parts separately have the product of dimers $\ket{D}$ with $\alpha=2i\sqrt{2}\Omega/\Delta\gamma$ as their steady state, which is therefore also the steady state of the total system. We note that this construction of the steady state out of the unidirectional steady state works only if the system has an even number of spins, that allows all of them to pair up in dimers. Only then the steady state of the cascaded parts to the left and to the right are compatible. If the number of spins is odd, the steady state of the unidirectional master equation is such that all spins pair up, in a product of dimers, except the last one, which goes to the well known mixed steady state of a single coherently driven two level system. This state is however not a steady state of the cascaded master equation to the opposite direction.

\subsection{Timescale for reaching the steady state}
\begin{figure}[t]
\includegraphics[width=0.48\textwidth]{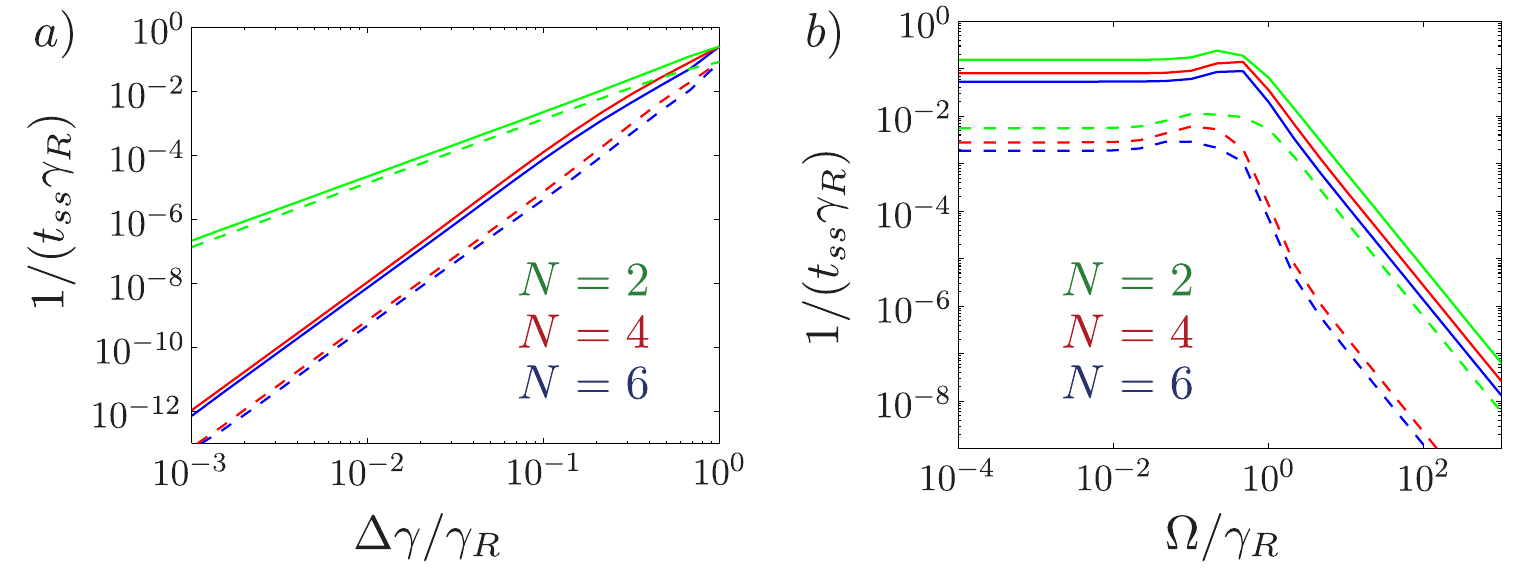}
\caption{Timescale $t_{\rm ss}$ to reach the unique stady state. (a) $t_{\rm ss}$ as a function of $\Delta\gamma$ for $\Omega/\gamma_R=0.5$ (solid lines) and $\Omega/\gamma_R=1$ (dashed lines). (b) $t_{\rm ss}$ as a function of $\Omega$ for $\Delta\gamma/\gamma_R=0.2$ (solid lines) and $\Delta\gamma/\gamma_R=0.8$ (dashed lines).}
\label{fig:gap}
\end{figure}
Quantitatively, the spin chain reaches the pure steady state on a timescale $t_{\rm ss}\equiv-1/{\rm Re}(\lambda_1)$, with $\lambda_1$ being the first nonzero eigenvalue of the total Liouvillian on the right hand side of Eq.\,(\ref{mastershort}). In Fig.\,\ref{fig:gap} we display $1/t_{\rm ss}$ on log-log scale as a function of $\Delta\gamma$ and $\Omega$, for different $N$. Leaving aside the special case of $N=2$, we can extract from Fig.~\ref{fig:gap}(a) the power law behaviour $t_{\rm ss}\sim \Delta\gamma^{-4}$.  This stems from the fact that for $\Delta\gamma=0$ the steady state is not unique. In this perfectly bidirectional case Eq.~\eqref{mastershort} has an additional symmetry and does not couple different Dicke manifolds. From Fig.~\ref{fig:gap}(b) we can extract that $t_{\rm ss}$ increases with the Rabi frequency $\Omega$ as $t_{\rm ss}\sim \Omega^2$ for $\Omega\gg\gamma_R$.

While the results presented in this work are for small number of spins, we also solved the master equation for larger system sizes using quantum trajectories \cite{Daley:2014hcSM}, demonstrating the formation of dimers and investigating the scaling of the dynamics with $N$ \cite{Pichler:2014abSM}.

\subsection{Photon scattering from Raman lasers does not limit the formation of dimers}

An experimental issue of using a Raman scheme to create SOC, is the heating of the quasi-BEC reservoir due to photon scattering \cite{Goldman:2013uqSM}. This can limit the lifetime of the quasi-BEC, specially in the case of light alkali atoms like Li or Na \cite{Goldman:2013uqSM}. However, we show here that in the case of a $^{87}$Rb SOC reservoir and for the moderate Raman strengths needed $\hbar\Omega_0<2E_0$, the photon scattering is negligible on the timescales related to the formation of dimers. For the SOC scheme with $^{87}$Rb considered in this work \cite{Lin:2011hnSM}, and assuming the lasers frequency is tuned exactly to the center on the fine-structure splitting $\Delta_{\rm FS}$, the Raman strength can be expressed as $\Omega_0\!=\!(\hbar\pi c^2\Gamma\Delta_{\rm FS}I)/(8\omega_{eg}^3\Delta_{\rm D2}^2)$ \cite{Goldman:2013uqSM}. Here $\omega_{eg}$ is the frequency of the $5S\rightarrow 5P$ transition, $\Gamma$ the corresponding natural linewidth, $I$ the lasers intensity, $\Delta_{\rm D2}=\Delta_{\rm FS}/2$ the detuning of the lasers from the D2 transition and $c$ the velocity of light. Under the same conditions, the photon scattering rate is given by $\Gamma_{\rm sc}\!=\!(3\pi c^2\Gamma^2 I)/(2\hbar\omega_{eg}^3\Delta_{\rm D2}^2)$ \cite{Grimm:1999wxSM}, implying the relation $\Gamma_{\rm sc}\!=\!12(\Gamma/\Delta_{\rm FS})\Omega_0\approx 10^{-5}\Omega_0$, in the case of $^{87}$Rb. As stated in the main text, we estimate $E_0/\hbar\approx 2\pi\cdot 3.5 {\rm kHz}$, and thus the maximum Raman coupling needed in our setup is on the order of $\Omega_0\sim 2E_0/\hbar\approx 2\pi\cdot 7{\rm kHz}$. As a result, the scattering rates will always be $\Gamma_{\rm sc}\lesssim 2\pi\cdot 7\cdot 10^{-2}{\rm Hz}$ and the scattering lifetimes $\tau\equiv 2\pi/\Gamma_{\rm sc}\gtrsim 14 {\rm s}$. For instance, in the case of a highly unidirectional reservoir with $\gamma_L/\gamma_R\sim 10^{-3}$, a system of 30 spins reaches the dimerized steady state on a timescale $t_{\rm ss}\sim 300/\gamma_R\sim 0.5\,{\rm s}\ll \tau$, i.e. much before the heating due to photon scattering becomes appreciable.


\begin{thebibliography}{10}
\bibitem{footnote} $n$ can also be half-integer, as one can reabsorb the resulting staggered phase by $\sigma_j\rightarrow (-1)^j\sigma_j$ and $\Omega_j\rightarrow (-1)^j\Omega_j$.
\end{thebibliography}
\end{document}